\documentclass[10pt,twocolumn,journal]{IEEEtran}
\usepackage{amsmath, graphics,amssymb,epsfig,subfigure,color}
\usepackage{algorithm}
\usepackage{algorithmic}
\usepackage{float}
\usepackage{multirow}
\usepackage{cuted,flushend}
\usepackage{midfloat}

\newtheorem{theorem}{Theorem}

\newtheorem{lemma}{Lemma}

\newtheorem{corollary}{Corollary}

\begin{document}
\title{Space-Time Interference Alignment and Degrees of Freedom Regions for the MISO Broadcast Channel with Periodic CSI Feedback  \thanks{N. Lee and R. Heath were funded in part by the Army Research Labs, Grant W911NF1010420. This paper was presented in part  \cite{Namyoon}  at IEEE 50th Annual Allerton Conference on
Communication, Control, and Computing, Monticello, Oct. 2012.}}

\author{  Namyoon Lee and Robert W. Heath Jr.\bigskip
\\
\normalsize Wireless Networking and Communications Group \\ \normalsize Department of Electrical and Computer Engineering
\\ \normalsize The University of Texas at
Austin, Austin, TX 78712 USA\\
      { \normalsize E-mail~:~namyoon.lee@utexas.edu, rheath@utexas.edu} }

\maketitle
\begin{abstract}
This paper characterizes the degrees of freedom (DoF) regions for the multi-user vector broadcast channel with periodic channel state information (CSI) feedback. As a part of the characterization, a new transmission method called \emph{space-time interference alignment} is proposed, which exploits both the current and past CSI jointly. Using the proposed alignment technique, an inner bound of the sum-DoF region is characterized as a function of a normalized CSI feedback frequency, which measures CSI feedback speed compared to the speed of user's channel variations. One consequence of the result is that the achievable sum-DoF gain is improved significantly when a user sends back both current and outdated CSI compared to the case where the user sends back current CSI only. Then, a trade-off  between CSI feedback delay and the sum-DoF gain is characterized for the multi-user vector broadcast channel in terms of a normalized CSI feedback delay that measures CSI obsoleteness compared to channel coherence time. A crucial insight is that it is possible to achieve the optimal DoF gain if the feedback delay is less than a derived fraction of the channel coherence time. This precisely characterizes the intuition that a small delay should be negligible. 



\end{abstract}
 
%


\section{Introduction}

Channel state information at the transmitter (CSIT) is important for optimizing wireless system performance. In the
multiple-input-single-output (MISO) broadcast channel, CSIT allows the transmitter to send multiple data symbols to
different receivers simultaneously without creating mutual interference by using interference suppression techniques \cite{Spencer}-\cite{Caire}.
Prior work on the MISO broadcast channel focused on the CSIT uncertainty caused by limited rate feedback \cite{Jindal}-\cite{Namyoon_IFBC}. This prior work showed no degrees of freedom (DoF) are lost compared to the perfect CSIT case when the CSI feedback rate per user linearly increases with signal to noise ratio (SNR) in dB scale. Meanwhile, there is little work on the effect of CSIT uncertainty due to feedback delay. In particular, if a transmitter has past CSI that is independent of the current CSI, i.e., outdated CSIT, it was commonly believed that only unity DoF gain could be achieved, which is the same DoF gain in the absence of channel knowledge at the transmitter \cite{Huang}. Recently, it was shown in \cite{Maddah-Ali2} that outdated CSIT is helpful to obtain the DoF gains beyond that achieved by time-division-multiple-access (TDMA). The key idea from \cite{Maddah-Ali2} was
to exploit the perfect outdated CSIT as side-information so that the transmitter aligns inter-user interference between the past and the currently received signals. Motivated by the work in \cite{Maddah-Ali2}, extensions have been developed for other networks \cite{Maleki}-\cite{Tandon}. The DoF gain was studied for the interference channel and the X channel when only outdated CSI is used at the transmitter in \cite{Maleki}. The DoF was characterized for the two-user MIMO
interference channel with delayed CSI according to different
transmit-and-receive-antenna configurations in \cite{Ghasemi}. The
outdated CSI feedback framework was also considered for the two-hop interference channel in \cite{Vaze} and
\cite{Tandon}, where the DoF gain increased beyond that achieved by the TDMA method.

The common assumption of previous work \cite{Maddah-Ali2}-\cite{Tandon} is that the transmitter has delayed
CSI only. Depending on a feedback protocol and relative feedback delay compared to channel variations, however, it may be possible for the transmitter to have knowledge of outdated as well as current CSI. Therefore, one of the fundamental questions is how much gain can the system provide if the transmitter can have both current and outdated CSI? An answer for this question is addressed in the periodic CSI feedback framework in this paper. 


\subsection{ Two Different Models of Periodic CSI Feedback}
Unlike aperiodic CSI feedback where each user opportunistically sends back CSI whenever the channel condition is changed \cite{Tang2}, periodic CSI feedback allows the transmitter to receive CSI periodically from mobile users through dedicated feedback channels. This is suitable for current cellular standards such as 3GPP LTE \cite{Andrews2} because it is implementable using a control signal with lower overhead. In such a periodic CSI feedback protocol, two different models can be considered depending on the dominant factor that causes delayed CSIT.


\begin{figure*}
\centering
\includegraphics[width=7in]{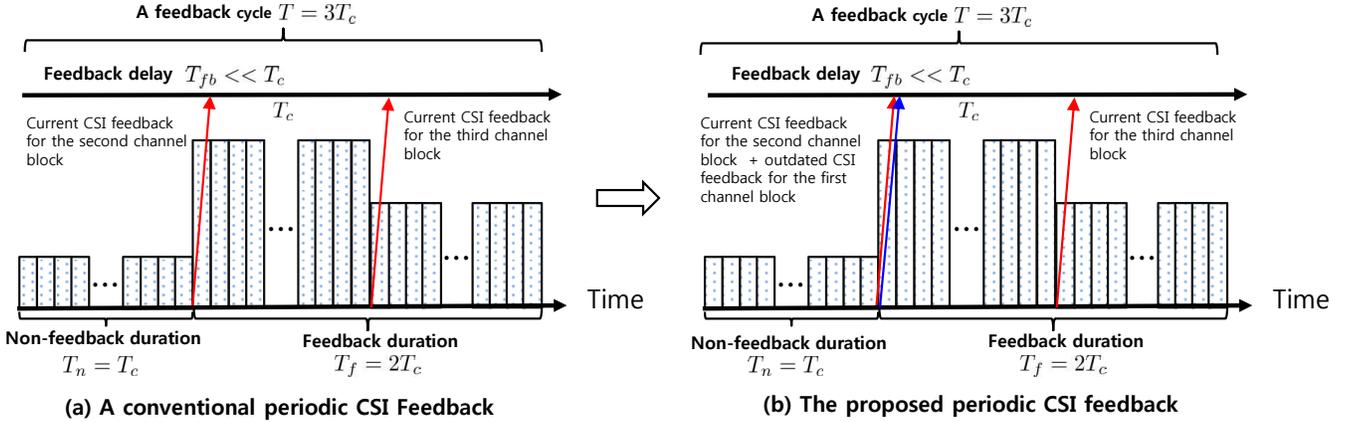}
\caption{Motivation of the proposed periodic CSI feedback model.  } \label{fig:motivation}
\end{figure*}

\begin{enumerate}
\item \textbf {Feedback frequency limited model (Model 1):} Consider a block fading channel where the channel values are constant for the channel coherence time $T_{\textrm{c}}$ and the channels change independently between blocks. Further, assume that feedback delay $T_{\textrm{fb}}$ is much less than the channel coherence time $T_{\textrm{c}}$. In such a channel, consider the case where a transmitter cannot track all channel changes of users because each user has a limited CSI feedback opportunity given by the periodic feedback protocol. In other words, the CSI feedback frequency is not high enough to continually track the user's channel variations. Consider the example illustrated in Fig. \ref{fig:motivation}-(a). If a user is able to use the feedback channel over the feedback duration $T_{\textrm{f}}=2T_{\textrm{c}}$ among the overall feedback period $T=3T_{\textrm{c}}$, i.e., $\omega=\frac{T_{\textrm{f}}}{T}=\frac{2}{3}$, the transmitter may have current CSI for two-thirds of the channel variations, assuming that the CSI feedback delay is much smaller than the channel coherence time. Instead of sending back currently estimated CSI only, however, the transmitter can acquire more channel knowledge, provided that each user sends back a set of CSI including outdated and current CSI during the feedback durations. For example, each user may send back one outdated and two current channel state estimates to the transmitter during feedback durations as shown in Fig. \ref{fig:motivation}-(b). In this model, CSIT is outdated due to the lower feedback frequency rather than the feedback link delay. Hence, it is possible to consider an equivalent CSI feedback model in a fast fading channel without considering feedback delay to focus on the effect of CSI feedback frequency as shown in Fig. \ref{fig:model}-(a). The equivalence comes from the fact that both models allow for the transmitter to use two outdated and one current observation of CSI periodically. In this model, we establish how the CSI feedback frequency affects the sum-DoF gain in this fast-fading channel configuration.


\begin{figure*}
\centering
\includegraphics[width=7in]{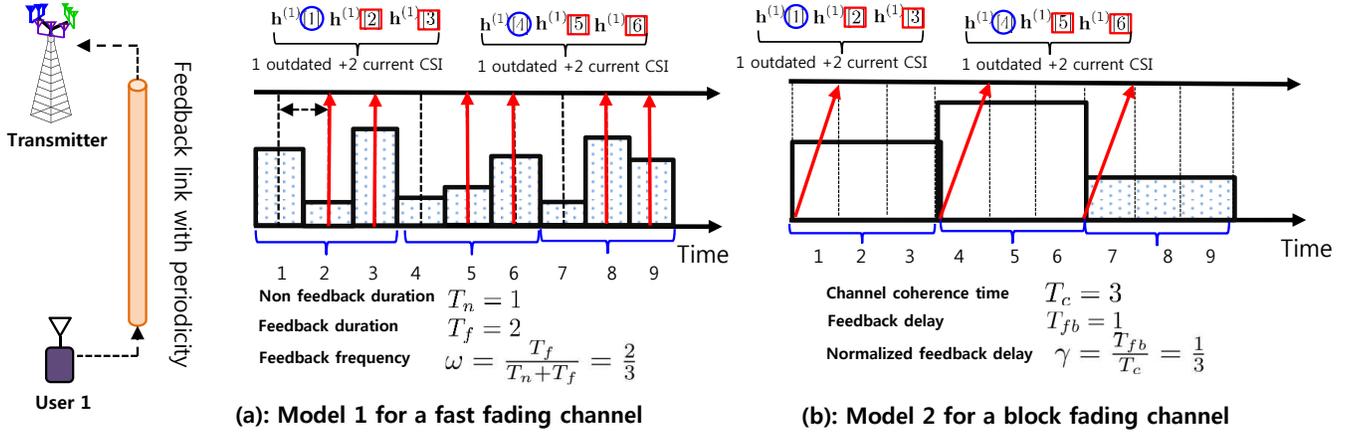}
\caption{Periodic CSI feedback models for both fast and block fading channels. In Model 1, outdated CSIT is due to the lower CSI feedback frequency than channel variations.  Meanwhile, outdated CSIT comes from the CSI feedback delay mainly in Model 2.} \label{fig:model}
\end{figure*}

\item  \textbf {Feedback link delay limited model (Model 2):} We also consider a different CSI feedback scenario in which the feedback link delay is the dominant factor that causes outdated CSIT. Consider a block fading channel where the transmitter can track all channel variations for each user, assuming that the feedback interval $T$ is equal to channel coherence time $T_{\textrm{c}}$ for simplicity. Further, the feedback delay $T_{\textrm{fb}}$ is not negligible compared to the channel coherence time. In this scenario, outdated CSIT is no longer affected by the feedback frequency because the transmitter can track the channel variations over all channel blocks. Rather, the relative difference between the CSI feedback delay and the channel coherence time causes the transmitter have the delayed CSI during a fraction of channel coherence time. For instance, when the feedback delay is one-third of channel coherence time, only delayed CSIT is available during one-third of channel coherence time, but both delayed and current CSIT is available in the remaining time as illustrated in Fig \ref{fig:model}-(b). 

\end{enumerate}

\subsection{Contribution}

\begin{figure}
\centering
\includegraphics[width=3.3in]{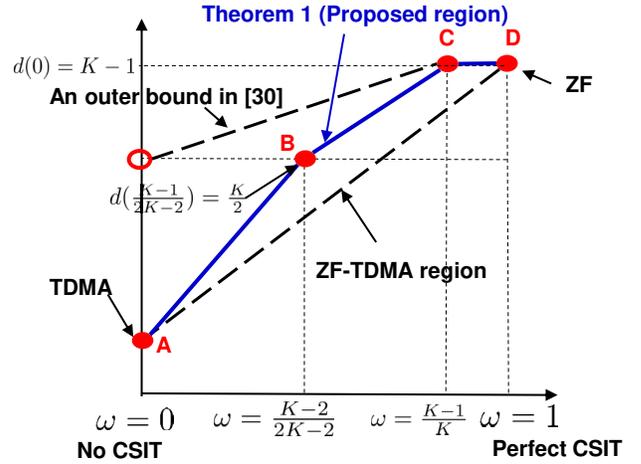}
\caption{An achievable sum-DoF region for the $K$-user
$(K-1)\times 1$ MISO broadcast channel with CSI feedback frequency limited model. The achievability of point B and C in the region is derived by the proposed STIA.} \label{fig:theorem1}
\end{figure}

In this paper, we consider the MISO broadcast channel where a transmitter with $N_\textrm{t}=K-1$
antennas supports $K$ users with a single antenna. The main contribution of this paper is to show that joint use of outdated and current CSIT improves significantly the sum-DoF of the vector broadcast channel in both periodic CSI feedback models. For the feedback frequency limited model, we provide a characterization of the achievable sum-DoF region as a function of CSI feedback frequency $\omega=\frac{T_{\textrm{f}}}{T_{\textrm{f}}+T_{\textrm{n}}}$. As illustrated in Fig. \ref{fig:theorem1}, we show that it is possible to increase the achievable sum-DoF region substantially from the additional outdated CSI feedback over all the range $\omega\in (0,1)$ compared to the case where each user sends back current CSI only at the feedback instant. One notable point is that when $\omega \geq \frac{K-1}{K}$, it is possible to achieve the optimal $\min\{N_\textrm{t},K\}=K-1$ DoF gain (cut-set bound) without knowledge of all current CSIT, i.e., $\omega<1$. This result is appealing because it was believed that the cut-set bound can be achieved when the transmitter only has current CSIT over all channel uses. Our result, however, reveals the fact that current CSIT over all channel uses is not necessarily required, and joint use of outdated and current CSIT allows the optimal DoF gain to be achieved.

For the feedback link delay limited model, we also characterize a trade-off region between CSI feedback delay and the sum-DoF gain for the multi-user vector broadcast channel by the proposed STIA algorithm and leveraging results in [11]. From the proposed region, we show that the joint use of outdated and current CSIT improves significantly the sum-DoF gain compared to that achieved by the separate exploitation of outdated and current CSIT. One consequence of the derived region is that not too delayed CSIT achieves the optimal DoF for the system. This is because the proposed trade-off region achieves the cut-set bound for the range of $0\leq \gamma \leq \frac{1}{K}$. Further, we demonstrate that the proposed trade-off region is optimal for the three-user $2\times 1$ vector broadcast channel using a recently derived converse result [30].

 To show these results, we propose a new transmission method called space-time interference alignment (STIA). The key idea of STIA is to design the transmit beamforming matrix by using both outdated and current CSI jointly so that multiple inter-user interference signals between the past observed and the currently observed are aligned in both space and time domains. Therefore, STIA is a generalized version of the transmission method developed in \cite{Maddah-Ali2} for using both outdated and current CSI.


%

In terms of related work, new transmission methods combining MAT in \cite{Maddah-Ali2} and partial zero-forcing (ZF) using both current and outdated CSI were developed for the two-user vector broadcast channel in \cite{Yang} and \cite{Gou}. The main difference lies in the system model. In  \cite{Yang} and \cite{Gou}, the imperfect current CSI estimated from the temporal channel correlations is used in the transmission algorithms. Meanwhile, our transmission algorithm exploits perfect current CSI. The problem of exploiting perfect outdated and current CSIT has been studied for the two-user vector broadcast channel \cite{Tandon_alt} to obtain the DoF gain, which is more closely related to our work. The limitation of work in \cite{Tandon_alt} is that it considers a simple two-user vector broadcast channel.


\begin{figure}
\centering
\includegraphics[width=3.5in]{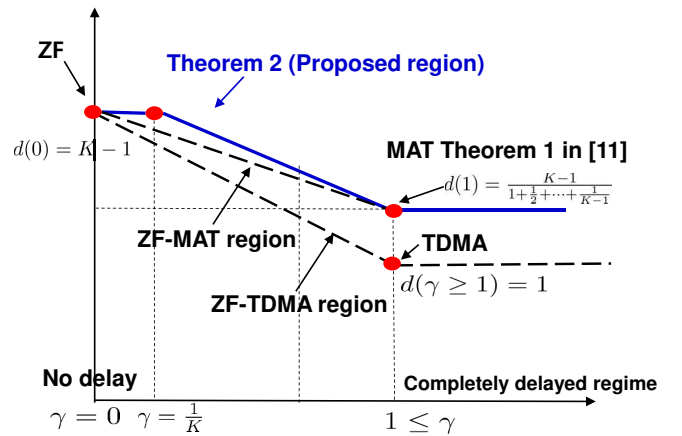}
\caption{A CSI feedback delay-DoF gain trade-off region for the $K$-user $(K-1)\times 1$ MISO broadcast channel with CSI feedback link delay limited model.} \label{fig:theorem2}
\end{figure}

This paper is organized as follows. In Section II,
the system model and periodic CSI feedback models are described. The key concept of the proposed
transmission method is explained by giving an illustrative
example in Section III. In Section IV, the achievable sum-DoF region for the $K$-user MISO broadcast channel with the feedback frequency limited model (Model 1) is characterized. In Section V, a CSI feedback delay and DoF gain trade-off is studied with the feedback link delay limited model (Model 2). In Section VI, the proposed algorithm is compared with the prior work. The paper is concluded in Section VII.

Throughout this paper, transpose, conjugate transpose, inverse, and
trace of a vector ${\bf x}$ are represented by ${\bf x}^{T}$ ,
${\bf x}^{*}$, ${\bf x}^{-1}$ and $\textmd{Tr}\left({\bf
x}\right)$, respectively. In addition, $\mathbb{C}$, $\mathbb{Z}^+$, and $\mathcal{CN}(0,1)$ a complex value, a positive integer value and a complex gaussian random variable with zero mean and unit variance.

\section{System Model}
\subsection{Singal Model}
Consider a $K$-user MISO broadcast channel where a transmitter with $N_\textrm{t}=K-1$ multiple antennas sends independent
messages to a receiver with a single antenna. The input-output
relationship at the $n$-th channel use is given by
\begin{align}
{y}^{(k)}[n]= {\bf h}^{(k)^T}[n]{\bf x}[n] + z^{(k)}[n],
\end{align}
where ${\bf x}[n]\in \mathbb{C}^{N_\textrm{t}\times 1}$ denotes the signal sent by the transmitter, ${{\bf
h}^{(k)}}[n]=\left[{h}^{(k)}_1[n], \ldots,{h}^{(k)}_{N_\textrm{t}}[n]\right]^T$ represents the
channel vector from the transmitter to user $k$; and $z^{(k)}[n]\sim
\mathcal{CN}(0,1)$ for $k\in\{1,2,\ldots,K\}$. We assume that the
transmit power at the transmitter satisfies an average constraint
$\mathbb{E}\left[\textmd{Tr}\left({\bf x}[n]{{\bf
x}}^{*}[n]\right)\right] \leq P$. We assume that all the entries of
channel vector ${\bf h}^{(k)}[n]$ are drawn from a continuous
distribution and the absolute value of all the channel coefficients is bounded between a nonzero minimum value and a finite maximum value to avoid the degenerate cases where channel values are equal to zero or infinity. Each receiver has a perfect estimate of its CSI, i.e. has perfect CSIR.

\subsection{Periodic CSI Feedback Model}
\subsubsection{Model 1 (CSI feedback frequency limited model)}

A feedback cycle ($T\in \mathbb{Z}^+$ time slots) consists of two sub durations: a non-feedback duration ($T_{\textrm{n}}$ time slots) and a CSI feedback duration ($T_{\textrm{f}}$ time slot). During the
non-feedback duration, each user cannot send back CSI to the
transmitter at all. Starting with the $T_{\textrm{n}}+1$ time slot, each user sends it back to the transmitter until $T_{\textrm{n}}+T_{\textrm{f}}$ time slots. In the
first time slot in the feedback duration, the user sends back both current CSI and the estimated CSI obtained from the previous non-feedback durations. From time slot $n=T_{\textrm{n}}+2$ to
$n=T_{\textrm{n}}+T_{\textrm{f}}$, only current CSI is provided to the transmitter
$\left\{{\bf h}^{(k)}[n]\right\}$. For example, user $k$ sends back $T_{\textrm{n}}-1$ outdated CSI $\left\{{\bf
h}^{(k)}[1],{\bf h}^{(k)}[2], \ldots,{\bf h}^{(k)}[T_{\textrm{n}}]\right\}$ and one current CSI, $\left\{{\bf h}^{(k)}[T_{\textrm{n}}\!+\!1]\right\}$ through the
feedback link at $T_{\textrm{n}}\!+\!1$ time slot. From time slot
$t=T_{\textrm{n}}\!+\!2$ to $t=T_{\textrm{n}}\!+\!T_{\textrm{f}}$, the only current CSI is sent back  to the
transmitter $\left\{{\bf h}^{(k)}[t]\right\}$. Since the channel
values are independently changed at every time slot, the transmitter
is able to obtain $T_{\textrm{n}}-1$ outdated or obsolete CSI $\left\{{\bf
h}^{(k)}[1],{\bf h}^{(k)}[2], \ldots,{\bf h}^{(k)}[T_{\textrm{n}}]\right\}$
and $T_{\textrm{f}}$ current CSI $\left\{{\bf h}^{(k)}[T_{\textrm{n}}\!+\!1],\ldots,{\bf h}^{(k)}[T_{\textrm{n}}\!+\!T_{\textrm{f}}]\right\}$ within a
feedback cycle $T=T_{\textrm{n}}+T_{\textrm{f}}$, where $k\in\{1,2,\ldots,K\}$. The CSI is provided
through a noiseless and delay-free feedback link at
regularly-spaced time intervals. 

In this model, we define a parameter that measures the normalized feedback speed, i.e., feedback frequency, which is
\begin{align}
\omega=\frac{T_{\textrm{f}}}{T_{\textrm{n}}+T_{\textrm{f}}}.
\end{align}
For example, if $\omega=1$, then the transmitter obtains instantaneous CSI over all time slots because the feedback is frequent enough to track all channel variations, i.e., the perfect CSIT case. Alternatively, if
$\omega=0$, then the transmitter cannot obtain any CSI from the receivers, i.e., no CSIT case.

\subsubsection{Model 2 (CSI feedback link delay limited model) }
We consider an ideal block fading channel where the channel values are constant for the channel coherence time $T_{\textrm{c}}$ and change independently between blocks. Each user sends backs CSI to the transmitter every $T_{\textrm{c}}$ time slots periodically where $T_{\textrm{c}}$ denotes channel coherence time. If we assume that the feedback delay time $T_{\textrm{fb}} $ is less than the channel coherence time, i.e., $T_{\textrm{fb}}<T_{\textrm{c}}$, then the transmitter learns the CSI $T_{\textrm{fb}}$ time slot after it was sent by the user. For example, if a user sends back CSI at time slot $n$, the transmitter has CSI at time slot $n+T_{\textrm{fb}}$ in our model.

Let us define a parameter for the ratio between the CSI feedback delay and the channel coherence time. We call this the normalized CSI feedback delay:
\begin{align}
\gamma=\frac{T_{\textrm{fb}}}{T_{\textrm{c}}}.
\end{align}
We refer to the case where $\gamma \geq 1$ as the completely outdated CSI regime as considered in \cite{Maddah-Ali2}. In this case, only completely outdated CSI  is available at the transmitter. We refer to the case where $\gamma=0$ as the current CSI point. Since there is no CSI feedback delay, the transmitter can use the current CSI in each slot. As depicted in Fig. 1-(b), if $\gamma=\frac{1}{3}$, the transmitter is able to exploit instantaneous CSI over two-thirds of the channel coherence time and outdated CSI for the previous channel blocks.

\subsection{Sum-DoF Region with Periodic CSI Feedabck}
Since the achievable data rate of the users depends on the parameters $\omega$ or $\gamma$ and SNR, it can be expressed as a function of $\lambda \in\{\omega,\gamma\}$ and $\textrm{SNR}$. Using this notion, for codewords spanning $n$ channel uses, a rate of user $k$
$R^{(k)}(\lambda,\textrm{SNR})=\frac{\log|m^{(k)}(\lambda,\textrm{SNR})|}{n}$ is achievable if the
probability of error for the message $m^{(k)}$ approaches zero as
$n\rightarrow \infty$. The sum-DoF characterizing the high SNR
behavior of the achievable rate region is defined as
\begin{align}
d(\lambda)=\sum_{k=1}^{K}d^{(k)}(\lambda)  =\lim_{\textrm{SNR}\rightarrow \infty}
\frac{\sum_{k=1}^{K}R^{(k)}(\lambda,\textrm{SNR})}{\log(\textrm{SNR})}.
\end{align}

\section{Space-Time Interference Alignment}
Before proving our main results, we explain the space-time interference alignment (STIA) signaling structure. The periodic CSI feedback protocol motivates us to develop the STIA because it exploits current and outdated CSI jointly. In this section, we focus on the special case of $K=3$, $N_\textrm{t}=2$, and $\omega=\frac{2}{3}$ to provide an intuition why joint use of both CSI is useful. We generalize this construction to a more general set of parameters in the sequel.

\subsection{Precoding and Decoding Structure}
In this section, we will show that $2$ of DoF (outer bound) is achieved for the $3$-user $2\times 1$ vector broadcast channel when $\omega=\frac{2}{3}$ ($T_{\textrm{n}}=1$ and $T_{\textrm{f}}=2$) under the assumptions of Model 1 as depicted Fig. 2-(a). The proposed STIA algorithm consists of two phases. The first phase involves sending multiple symbols per user to create an interference pattern without CSI knowledge at the transmitter. The second phase involves sending linear combinations of the same symbols to recreate the same interference patterns using both current and outdated CSI. Using the fact that each user receives the same interference patterns during the two phases, it cancels the interference signals to obtain the desired symbols in the decoding procedure.

\subsubsection{Phase One (Obtain the Interference Pattern)} This phase consists of one time slot.
During phase one, the transmitter does not have any CSI knowledge according to Model I because each user cannot feedback CSI during the non-feedback duration, i.e., $T_{\textrm{n}}=1$. In time slot 1, the transmitter sends six independent
symbols where $s^{(1)}_1$ and
$s^{(1)}_2$ intended for user 1, $s^{(2)}_1$ and
$s^{(2)}_2$ intended for user 2, and $s^{(3)}_1$ and
$s^{(3)}_2$ intended for user 3 without precoding as
\begin{align}
{\bf x}[1]= \sum_{k=1}^{3}{\bf s}^{(k)},
\end{align}
where ${\bf
s}^{(k)}=\left[s^{(k)}_1, s^{(k)}_2\right]^{T}$. Neglecting noise at the receiver, the observation at each receiver is a function of three linear combinations:
\begin{align}
y^{(1)}[1]&= L^{(1,1)}[1] + L^{(1,2)}[1]+ L^{(1,3)}[1], \\
y^{(2)}[1]&= L^{(2,1)}[1] + L^{(2,2)}[1]+ L^{(2,3)}[1],\\
y^{(3)}[1]&= L^{(3,1)}[1] + L^{(3,2)}[1]+ L^{(3,3)}[1],
\end{align}
where $L^{(k,i)}[n]$ denotes a linear combination seen by user $k$ for the transmitted symbols for user $i$ at the $n$-th time slot:
 \begin{align}
L^{(1,1)}[1]&=h^{(1)}_1[1]s^{(1)}_1+h^{(1)}_2[1]s^{(1)}_2, \nonumber \\
L^{(1,2)}[1]&=h^{(1)}_1[1]s^{(2)}_1+h^{(1)}_2[1]s^{(2)}_2,
\nonumber \\
L^{(1,3)}[1]&=h^{(1)}_1[1]s^{(3)}_1+h^{(1)}_2[1]s^{(3)}_2,
\nonumber \\
L^{(2,1)}[1]&=h^{(2)}_1[1]s^{(1)}_1+h^{(2)}_2[1]s^{(1)}_2,\nonumber \\
 L^{(2,2)}[1]&=h^{(2)}_1[1]s^{(2)}_1+h^{(2)}_2[1]s^{(2)}_2, \nonumber \\
L^{(2,3)}[1]&=h^{(2)}_1[1]s^{(3)}_1+h^{(2)}_2[1]s^{(3)}_2,\nonumber \\
L^{(3,1)}[1]&=h^{(3)}_1[1]s^{(1)}_1+h^{(3)}_2[1]s^{(1)}_2,\nonumber \\
L^{(3,2)}[1]&=h^{(3)}_1[1]s^{(2)}_1+h^{(3)}_2[1]s^{(2)}_2,
\nonumber \\ 
\textrm{and} \quad L^{(3,3)}[1]&=h^{(3)}_1[1]s^{(3)}_1+h^{(3)}_2[1]s^{(3)}_2.\nonumber 
\end{align}
In summary, during phase one, each receiver obtains a linear combination of one desired equation $L^{(i,i)}[1]$ and two interference equations $L^{(i,j)}[1]$ for $i\neq j$. 

\subsubsection{Phase Two (Generate the Same Interference Pattern)} 
The second phase uses two time slots, i.e. $n \in \{2,3\}$. In this phase, the transmitter has knowledge of both current and outdated CSI thanks to feedback. For example, during time slot 2, the transmitter has outdated CSI for the first time slot  and current CSI for the second time slot. Using this information, in time slot 2 and time slot 3, the transmitter sends simultaneously two symbols for the dedicated users by using linear beamforming as
\begin{align}
{\bf x}[n]= \sum_{k=1}^{3}{\bf V}^{(k)}[n]{\bf s}^{(k)},
\end{align}
where ${\bf V}^{(k)}[n]\in \mathbb{C}^{2\times 2}$ denotes the
beamforming matrix used for carrying the same symbol vector ${\bf
s}^{(k)}=\left[s^{(k)}_1, s^{(k)}_2\right]^T$ at time slot $n$, where $n\in\{2,3\}$ and $k\in\{1,2,3\}$. The main idea for designing precoding matrix ${\bf V}^{(k)}[n]$ is to allow all the receivers observe the same linear combination for interference signals received during time slot 1 by exploiting current and outdated CSI. For instance,
user 2 and user 3 received the interference signals in the form of
$L^{(2,1)}[1]=h^{(2)}_1[1]s^{(1)}_1+h^{(2)}_2[1]s^{(1)}_2$ and
$L^{(3,1)}[1]=h^{(3)}_1[1]s^{(1)}_1+h^{(3)}_2[1]s^{(1)}_2$. Therefore, to deliver the same linear combination for the undesired symbols to user 2 and user 3 repeatedly during $n\in\{2,3\}$, the transmitter constructs the beamforming
matrix carrying symbols, $s^{(1)}_1$ and $s^{(1)}_2$ to satisfy
\begin{align}
\left[%
\begin{array}{c}
  {{\bf h}^{(2)T}}[n] \\
  {{\bf h}^{(3)T}}[n] \\
\end{array}%
\right]{\bf V}^{(1)}[n]=\left[%
\begin{array}{c}
  {{\bf h}^{(2)T}}[1] \\
  {{\bf h}^{(3)T}}[1] \\
\end{array}%
\right].
\end{align} 
Similarly, to ensure the interfering users receive the same linear
combination of the undesired symbols, which is linearly dependent
(aligned) with the previously overheard equation during time slot 1, the beamforming matrices carrying data symbols for user 2 and user 3 are constructed to satisfy the space-time inter-user interference alignment conditions:
\begin{align}
\left[%
\begin{array}{c}
  {{\bf h}^{(1)T}}[n] \\
  {{\bf h}^{(3)T}}[n] \\
\end{array}%
\right]{\bf V}^{(2)}[n]=\left[%
\begin{array}{c}
  {{\bf h}^{(1)T}}[1] \\
  {{\bf h}^{(3)T}}[1] \\
\end{array}%
\right],
\end{align}
and
\begin{align}
\left[%
\begin{array}{c}
  {{\bf h}^{(1)T}}[n] \\
  {{\bf h}^{(2)T}}[n] \\
\end{array}%
\right]{\bf V}^{(3)}[n]=\left[%
\begin{array}{c}
  {{\bf h}^{(1)T}}[1] \\
  {{\bf h}^{(2)T}}[1] \\
\end{array}%
\right].
\end{align}
Since we assume that the channel coefficients are drawn from a continuous distribution, matrix inversion is guaranteed with high probability. Therefore, it is possible to construct transmit
beamforming matrices ${\bf V}^{(1)}[n]$, ${\bf V}^{(2)}[n]$ and ${\bf
V}^{(3)}[n]$ as
\begin{align}
{\bf V}^{(1)}[n]&=\left[%
\begin{array}{c}
  {{\bf h}^{(2)T}}[n] \\
  {{\bf h}^{(3)T}}[n] \\
\end{array}%
\right]^{-1}\left[%
\begin{array}{c}
  {{\bf h}^{(2)T}}[1] \\
  {{\bf h}^{(3)T}}[1] \\
\end{array}%
\right], \\
{\bf V}^{(2)}[n]&=\left[%
\begin{array}{c}
  {{\bf h}^{(1)T}}[n] \\
  {{\bf h}^{(3)T}}[n] \\
\end{array}%
\right]^{-1}\left[%
\begin{array}{c}
  {{\bf h}^{(1)T}}[1] \\
  {{\bf h}^{(3)T}}[1] \\
\end{array}%
\right],
\end{align}
and
\begin{align}
{\bf V}^{(3)}[n]&=\left[%
\begin{array}{c}
  {{\bf h}^{(1)T}}[n] \\
  {{\bf h}^{(2)T}}[n] \\
\end{array}%
\right]^{-1}\left[%
\begin{array}{c}
  {{\bf h}^{(1)T}}[1] \\
  {{\bf h}^{(2)T}}[1] \\
\end{array}%
\right].
\end{align}
If we denote ${{\bf \tilde{h}}^{(1)T}}[n]={{\bf h}^{(1)T}}[n]{\bf V}^{(1)}[n]$ and $L^{(1,1)}[n]={{\bf \tilde{h}}^{(1)T}}[n]{\bf s}^{(1)}$ for $n=2,3$, at time slot 2 and time slot 3, the received signals at user 1 are given by
\begin{align}
{y}^{(1)}[2]&= \sum_{k=1}^{3}{{\bf h}^{(1)T}}[2]{\bf V}^{(k)}[2]{\bf s}^{(k)}\nonumber \\
&={{\bf h}^{(1)T}}\![2]{\bf V}^{(1)}[2]{\bf s}^{(1)}+{{\bf h}^{(1)T}}[2]{\bf V}^{(2)}\![2]{\bf s}^{(2)}\nonumber \\ &+{{\bf h}^{(1)T}}[2]{\bf V}^{(3)}[2]{\bf s}^{(3)} \nonumber \\
&={{\bf \tilde{h}}^{(1)T}}[2]{\bf s}^{(1)}+{{\bf {h}}^{(1)T}}[1]{\bf s}^{(2)}+{{\bf {h}}^{(1)T}}[1]{\bf s}^{(3)}  \nonumber \\
&=L^{(1,1)}[2] + L^{(1,2)}[1]+ L^{(1,3)}[1], \label{eq:STIA_ex1}\\
{y}^{(1)}[3]&= \sum_{k=1}^{3}{{\bf h}^{(1)T}}[3]{\bf V}^{(k)}[3]{\bf s}^{(k)}\nonumber \\
&={{\bf h}^{(1)T}}[3]{\bf V}^{(1)}[3]{\bf s}^{(1)}+{{\bf h}^{(1)T}}\![3]{\bf V}^{(2)}\![3]{\bf s}^{(2)}\nonumber \\ &+{{\bf h}^{(1)T}}[3]{\bf V}^{(3)}\![3]{\bf s}^{(3)} \nonumber \\
&={{\bf \tilde{h}}^{(1)T}}[3]{\bf s}^{(1)}+{{\bf {h}}^{(1)T}}[1]{\bf s}^{(2)}+{{\bf {h}}^{(1)T}}[1]{\bf s}^{(3)}  \nonumber \\
&=L^{(1,1)}[3] + L^{(1,2)}[1]+ L^{(1,3)}[1].\label{eq:STIA_ex2}
\end{align}

If we denote ${{\bf \tilde{h}}^{(2)T}}[n]={{\bf h}^{(2)T}}[n]{\bf V}^{(2)}[n]$ and $L^{(2,2)}[n]={{\bf \tilde{h}}^{(2)T}}[n]{\bf s}^{(2)}$ for $n=2, 3$, the received signals at user 2 during time slot 2 and 3 are given by
\begin{align}
{y}^{(2)}[2]&=\sum_{k=1}^{3}{{\bf h}^{(2)T}}[2]{\bf V}^{(k)}[2]{\bf s}^{(k)}\nonumber \\
&={{\bf h}^{(2)T}}[2]{\bf V}^{(1)}[2]{\bf s}^{(1)}+{{\bf h}^{(2)T}}[2]{\bf V}^{(2)}[2]{\bf s}^{(2)}\nonumber \\ &+{{\bf h}^{(2)T}}[2]{\bf V}^{(3)}[2]{\bf s}^{(3)} \nonumber \\
&={{\bf {h}}^{(2)T}}[1]{\bf s}^{(1)}+{{\bf {\tilde h}}^{(2)}}[2]{\bf s}^{(2)}+{{\bf {h}}^{(2)T}}[1]{\bf s}^{(3)}  \nonumber \\
&=L^{(2,1)}[1] + L^{(2,2)}[2]+ L^{(2,3)}[1],\\
{y}^{(2)}[3]&= \sum_{k=1}^{3}{{\bf h}^{(2)T}}[3]{\bf V}^{(k)}[3]{\bf s}^{(k)}\nonumber \\
&={{\bf h}^{(2)T}}[3]{\bf V}^{(1)}\![3]{\bf s}^{(1)}+{{\bf h}^{(2)T}}[3]{\bf V}^{(2)}[3]{\bf s}^{(2)}\nonumber \\ &+{{\bf h}^{(2)T}}[3]{\bf V}^{(3)}[3]{\bf s}^{(3)} \nonumber \\
&={{\bf {h}}^{(2)T}}[1]{\bf s}^{(1)}+{{\bf {\tilde h}}^{(2)T}}[3]{\bf s}^{(2)}+{{\bf {h}}^{(2)T}}[1]{\bf s}^{(3)}  \nonumber \\
&=L^{(2,1)}[1] + L^{(2,2)}[3]+ L^{(2,3)}[1].
\end{align}

Finally, for user 3, if we denote ${{\bf \tilde{h}}^{(3)T}}[n]={{\bf h}^{(3)T}}[n]{\bf V}^{(3)}[n]$ and $L^{(3,3)}[n]={{\bf \tilde{h}}^{(3)T}}[n]{\bf s}^{(3)}$ for $n=2,3$, the received signals at time slot 2 and 3 are given by
\begin{align}
{y}^{(3)}[2]&= \sum_{k=1}^{3}{{\bf h}^{(3)T}}[2]{\bf V}^{(k)}[2]{\bf s}^{(k)}\nonumber \\
&={{\bf h}^{(3)T}}[2]{\bf V}^{(1)}[2]{\bf s}^{(1)}+{{\bf h}^{(2)T}}[2]{\bf V}^{(2)}[2]{\bf s}^{(2)}\nonumber \\ &+{{\bf h}^{(3)T}}[2]{\bf V}^{(3)}[2]{\bf s}^{(3)} \nonumber \\
&={{\bf {h}}^{(3)T}}[1]{\bf s}^{(1)}+{{\bf { h}}^{(3)T}}[1]{\bf s}^{(2)}+{{\bf {\tilde h}}^{(3)T}}[2]{\bf s}^{(3)}  \nonumber \\
&=L^{(3,1)}[1] + L^{(3,2)}[1]+ L^{(3,3)}[2],
\end{align}
\begin{align}
{y}^{(3)}[3]&= \sum_{k=1}^{3}{{\bf h}^{(3)T}}[3]{\bf V}^{(k)}[3]{\bf s}^{(k)}\nonumber \\
&={{\bf h}^{(3)T}}[3]{\bf V}^{(1)}[3]{\bf s}^{(1)}+{{\bf h}^{(3)T}}[3]{\bf V}^{(2)}[3]{\bf s}^{(2)}\nonumber \\ &+{{\bf h}^{(3)T}}[3]{\bf V}^{(3)}[3]{\bf s}^{(3)} \nonumber \\
&={{\bf {h}}^{(3)T}}[1]{\bf s}^{(1)}+{{\bf { h}}^{(3)T}}[1]{\bf s}^{(2)}+{{\bf {\tilde h}}^{(3)T}}[3]{\bf s}^{(3)}  \nonumber \\
&=L^{(3,1)}[1] + L^{(3,2)}[1]+ L^{(3,3)}[3].
\end{align}

\subsubsection{Decoding}
To explain the decoding process, let us consider the received signals at user 1. In time slot 1, user 1 acquired knowledge of the interference signals in the form of $L^{(1,2)}[1]$ and $L^{(1,3)}[1]$. From phase two, user 1 received the same interference signals $L^{(1,2)}[1]$ and $L^{(1,3)}[1]$ at time slot 2 and 3 as shown in (\ref{eq:STIA_ex1}) and (\ref{eq:STIA_ex2}). Therefore, to decode the desired signals, the observation made during phase two are subtracted from the observation in phase one, resulting in the equations:
\begin{align}
y^{(1)}[1]\!-\!y^{(1)}[2]&=L^{(1,1)}[1] + L^{(1,2)}[1]+ L^{(1,3)}[1] \nonumber \\
&-L^{(1,1)}[2] - L^{(1,2)}[1] - L^{(1,3)}[1] \nonumber \\
&= L^{(1,1)}[1] - L^{(1,1)}[2]
\nonumber \\
&=\left({\bf h}^{(1)T}[1]-{\bf h}^{(1)T}[2]{\bf V}^{(1)}[2]\right){\bf s}^{(1)},
\end{align}
\begin{align}
y^{(1)}[1]\!-\!y^{(1)}[3]&=L^{(1,1)}[1] + L^{(1,2)}[1]+ L^{(1,3)}[1] \nonumber \\
&-L^{(1,1)}[3] - L^{(1,2)}[1] - L^{(1,3)}[1] \nonumber \\
&= L^{(1,1)}[1] - L^{(1,1)}[3]
\nonumber \\
&= \left({\bf h}^{(1)T}[1]-{\bf h}^{(1)T}[3]{\bf V}^{(1)}[3]\right){\bf s}^{(1)}.
\end{align}
After removing the interference signals, the effective
channel input-output relationship for user 1 during the three time slots is given 
by
\begin{align}
\left[\!\!%
\begin{array}{c}
  {y}^{(1)}[1]-{y}^{(1)}[2]\\
  {y}^{(1)}[1]-y^{(1)}[3] \\
\end{array}%
\!\!\right]\!\!=\!\!\underbrace{\left[%
\begin{array}{c}
  {\bf h}^{(1)T}[1]-{\bf h}^{(1)T}[2]{\bf V}^{(1)}[2] \\
 {\bf h}^{(1)T}[1]-{\bf h}^{(1)T}[3]{\bf V}^{(1)}[3]\\
\end{array}%
\right]}_{{\bf H}^{(1)}_{\textrm{eff}}}\left[\!\!%
\begin{array}{c}
  s^{(1)}_1 \\
  s^{(1)}_2 \\
\end{array}\!\!%
\right].  \label{eq:Decod}
\end{align}
 Since precoding matrix ${\bf V}^{(1)}[n]$ for $n=2,3$ was designed independently of channel ${\bf h}^{(1)T}[1]$, the elements of the effective channel vector observed at time slot 2 and 3, i.e., $\left[\tilde{h}^{(1)}_1[2],
\tilde{h}^{(1)}_2[2]\right]={{\bf h}^{(1)T}}[2]{\bf V}^{(1)}[2]$ and $\left[\tilde{h}^{(1)}_1[3],
\tilde{h}^{(1)}_2[3]\right]={{\bf h}^{(1)T}}[3]{\bf V}^{(1)}[3]$ are also statistically independent random variables. This implies that the three channel vectors, ${\bf h}^{(1)T}[1]$, ${{\bf h}^{(1)T}}[2]{\bf V}^{(1)}[2]$, and ${{\bf h}^{(1)T}}[3]{\bf V}^{(1)}[3]$ are linearly independent. Therefore, 
$\textrm{rank}\left({\bf H}^{(1)}_{\textrm{eff}}\right)=2$ with a probability of one because all channel elements are drawn from a continuous distribution. As a result, user 1 decodes two desired symbols within three time
slots. In the same way, user 2 and user 3 are able to retrieve a
linear combination of their desired symbols by removing the
interference signals and can use the same decoding method. Since the transmitter has delivered two
independent symbols for its intended user in three channel uses,
a total $d=\frac{6}{3}=2$ DoF are achieved.

\subsection{The Relationship between STIA and Wireless Index Coding}
 
In the above example, we showed that the proposed STIA achieves the optimal DoF for the 3-user $2\times 1$ vector broadcast channel without using current CSI over all channel time slots. We can interpret STIA from a wireless index coding perspective. The index coding problem was introduced in \cite{Birk} and has been studied in subsequent work \cite{Yossef}-\cite{Maleki2}. Suppose that a transmitter has a set of information messages $\mathcal{W}=\{W_1,W_2,\ldots,W_K\}$ for multiple receivers and each receiver wishes to receive a subset of $\mathcal{W}$ while knowing some another subset of $\mathcal{W}$ as side-information. The index coding problem is to design the best encoding strategy at the transmitter, which minimizes the number of transmissions while ensuring that all receivers can obtain the desired messages. To illustrate, consider the case where a transmitter with a single antenna serves two single antenna users. Suppose each user has the message for the other user as side-information, i.e., user 1 has $W_2$ and user 2 has $W_1$. In this case, an index coding method is to send the linear sum of two messages $W_1+W_2$ to both receivers. Since user 1 and 2 receive $y^{(1)}[1]=h^{(1)}_1[1](W_1+W_2)$ and $y^{(2)}[1]=h^{(2)}_1[1](W_1+W_2)$ ignoring the noise and assuming the flat fading channel, each receiver can extract the desired message by using side-information as $\hat{W}_1=y^{(1)}[1]/h^{(1)}_1[1] -W_2$ and $\hat{W}_2=y^{(2)}[1]/h^{(2)}_1[1] -W_1$. STIA mimics this index coding algorithm in a more general sense by using current and outdated CSI jointly. This is because, during phase one, each user acquires side-information in a form of a linear combination of all transmitted data symbols where the linear coefficients are created by a wireless channel. During the second phase, the transmitter constructs the precoding matrices by using outdated and current CSI so that each user can exploit the received interference equations in the first phase as side-information, which results in minimizing the number of transmissions; it leads to an increase in the DoF.

\subsection{Extensions of STIA Algorithm}
We proposed the STIA algorithm as a way to take advantage of current and past CSI in the vector broadcast channel. Now, we make some remarks about how the STIA algorithm can be modified and extended to deal with more complex networks and practical aspects. 

\textbf{Remark 1 (Extension into interference networks \cite{Namyoon_IC} )}: The
proposed algorithm is directly applicable to the $K$-user MISO
interference channel where each transmitter has $K-1$ antennas.
The reason is that the beamforming matrices used in the second phase are independently generated without knowledge of the other user's data symbols. In other words, data sharing among the different transmitters is not required to apply the STIA in the $K$-user MISO interference channel. For example, the $\frac{3}{2}$ of DoF gain is also achievable for the $3$-user MISO interference channel with the periodic feedback
model when $N_{\textrm{t}}=2$.

%

\textbf{Remark 2 (Transmission power constraint)}:
Since it is assumed that the transmitter sends a signal with large enough power in the DoF analysis, the beamforming solutions containing matrix inversion do not violate the transmit power constraint. In practice, however, the proposed STIA algorithm needs to be modified so that the power constraint is satisfied. This modification may incur the sum-rate performance loss but does not affect the DoF gain. The sum-rate maximization or mean square error minimization  problems of the STIA algorithm given transmit power constraint can be investigated in future work.

\textbf{Remark 3 (Effective channel estimation in multi-carrier systems)}:
In the STIA algorithm, each receiver requires the effective channel during phase two. For example, user $1$ has to know $\{ {\bf h}^{(1)T}[2]{\bf V}^{(1)}[2]  \}$ and $\{ {\bf h}^{(1)T}[3]{\bf V}^{(1)}[3]  \}$ when it decodes the desired signal as shown in (\ref{eq:Decod}). This effective channel knowledge may be acquired at the receiver if the transmitter sends this information as data to the user by spending other downlink channel resources, i.e., a feed-forward information. This feed-forward, however, is a source of overhead in the system. Assuming a multi-carrier system, effective channel knowledge at the receiver can be obtained without feed-forward processing. The main idea for the effective channel estimation at the receiver is to use a demodulation reference signal and an interference cancellation technique. 

Let us explain this by giving an example that uses a multi-carrier system, which allows channel coherence in the frequency domain to be exploited. Consider a multi-carrier system where the channel values are the same across channel coherence band $B_{\textrm{c}}$, i.e., ${\bf h}^{(1)}[t,f_j]={\bf h}^{(1)}[t]$ for $t \in\{2,3\}$ and $j\in \{1,2,\ldots, B_{\textrm{c}}\}$. When the transmitter sends the training signal (demodulation reference signal) ${\bf t}[2,f_j]$ by the same precoding method used for data signal as ${\bf \tilde{t}}[2,f_j]=\sum_{k=1}^{K}{\bf V}^{(k)}[2,f_j] {\bf t}[2,f_j]$ through subcarrier $f_j$ where $j\in\{1,2,\ldots,B_{\textrm{t}}\}$ and $B_{\textrm{t}} <B_{\textrm{c}}$. The received signal at time slot $2$ and subcarrier $f_j$ is given by
\begin{align}
y^{(1)}[2,f_j]&={\bf h}^{(1)T}[2,f_j]{\bf \tilde{t}}[2,f_j]\nonumber \\
&=\sum_{k=1}^{3}{\bf h}^{(1)T}[2,f_j]{\bf V}^{(k)}[2,f_j] {\bf t}[2,f_j]\nonumber \\
&= {\bf h}^{(1)T}[2,f_j]{\bf V}^{(1)}[2] {\bf t} [2,f_j]+ 2{\bf h}^{(1)T}[1,f_j]{\bf t}[2,f_j],
\end{align}
where the last equality comes from the fact that ${\bf h}^{(1)T}[1,f_j]={\bf h}^{(1)T}[2,f_j]{\bf V}^{(k)}[2,f_j]$ for $k\in\{2,3\}$ by the STIA algorithm. Note that  each receiver has knowledge of training signal ${\bf t}[2,f_j]$ and the channel at time slot 1 ${\bf h}^{(1)T}[1,f_j]$. Hence, user 1 can extract out the desired equation for the channel estimation at $f_j$ subcarrier as $y^{(1)}[2,f_j]-2{\bf h}^{(1)T}[1,f_j]{\bf t}[2,f_j]= {\bf h}^{(1)T}[2,f_j]{\bf V}^{(1)}[2,f_j] {\bf t}[2,f_j]$ where $j\in\{1,2,\ldots,B_{\textrm{t}}\}$. As an example, the receiver could estimate the effective channel $ {\bf \tilde{h}}^{(1)}=({\bf h}^{(1)T}[2,f_j]{\bf V}^{(1)}[2,f_j])^* $ by solving the equation,
\begin{align}
\left[\!\!\!%
\begin{array}{c}
  y^{(1)*}[2,f_1]\!-\!(2{\bf h}^{(1)T}[1,f_1]{\bf t}[2,f_1])^* \\
  \vdots \\
  y^{(1)*}[2,f_{B_{\textrm{t}}}]\!-\!(2{\bf h}^{(1)T}[1,f_{B_{\textrm{t}}}]{\bf t}[2,f_{B_{\textrm{t}}}])^* \\
\end{array}%
\!\!\!\!\right]&\!\!=\! \left[\!\!\!\!%
\begin{array}{c}
 {\bf t}^*[2,f_1]\\
 \vdots \\
 {\bf t}^*[2,f_{B_{\textrm{t}}}] \\
\end{array}%
\!\!\!\right]{\bf \tilde{h}}^{(1)}.
\end{align}
Since the size of effective channel $ {\bf \tilde{h}}^{(1)}$ is $2\times 1$, if the transmitter uses an independent training signal $ {\bf t}^*[2,f_j]$ over $B_{\textrm{t}} \geq 2$ subcarriers among total $B_{\textrm{c}}$ subcarriers, it is estimated reliably, if the noise effect is ignored. Therefore, the STIA algorithm can be implemented without forwarding the effective channel to the users. Further, the overhead required for channel estimation is the same as that required for a multiuser MISO system. This implies that the STIA algorithm does not require additional overhead for learning the effective channel in multi-carrier systems.

\section{An Inner Bound of Sum-DoF Region}

In this section, we characterize an achievable sum-DoF region as a function of feedback frequency for the $K$-user $(K-1) \times 1$ MISO broadcast channel. The main result is established in the following theorem.

\begin{theorem} \label{Theorem1}
For the $K$-user $(K-1) \times 1$ MISO broadcast channel with the periodic CSI
feedback (Model 1), the achievable sum-DoF region is characterized as a function of feedback frequency $\omega$,
\begin{align}
d(\omega) \!=\! \left\{
\!\!\!\begin{array}{l l}
  (K-1)\omega+1, \!\!\!\!& \!\!\!\!\quad \textrm{for} \quad  \!\!0 \leq \omega \leq \frac{K-2}{2(K-1)}, \\
  a(K)\omega+b(K), \!\!\!\!& \!\!\!\!\quad \textrm{for} \quad \!\!\frac{K-2}{2(K-1)} \leq \omega \leq \frac{K-1}{K},\\ 
  K-1 \!\!\!\!& \!\!\!\!\quad \textrm{for} \!\!\quad \frac{K-1}{K} \leq \omega \leq 1.\\
\end{array} \right. 
\end{align}
where $a(K)=\frac{K(K-1)(K-2)}{(K-1)(K-2)+K}$ and $b(K)=\frac{K(K-1)}{(K-1)(K-2)+K}$.
\end{theorem}

\proof We prove Theorem 1 by four different transmission methods, each of which achieves four different corner points of the region as shown in Fig. \ref{fig:theorem1}. 
Since time sharing can be used to achieve the lines connecting each point, we focus on characterizing the conner points.  

\subsection{Achievablity of Point A and D}
Achievability for the corner point A when $\omega=0$ is shown by TDMA. Also, achievability for the point D when $\omega=1$ is proven by ZF. When receivers do not send back CSI to the transmitter, i.e., $\omega=0$, then the TDMA method is used to achieve $d(0)=1$ \cite{Huang}. At the other extreme, when the transmitter has current CSI over all time slots at $\omega=1$, it is possible to achieve the cut-set bound $d(1)=\min\{N_\textrm{t},K\}=K-1$ DoF by the ZF method.

\subsection{Achievability of Point B}

Let us consider $\omega=\frac{K-2}{2K-2}$, where the non-CSI feedback duration $T_{\textrm{n}}$ has $K$ time slots and the CSI feedback duration $T_{\textrm{f}}$ has  $K-2$ time slots. For this case, we show that $d\left(\frac{K-2}{2K-2}\right)=\frac{K(K-1)}{2K-2}=\frac{K}{2}$ DoF are achievable by STIA using $K$ time slots in the first phase and $K-2$ time slots in the second phase.

\subsubsection{Phase One} The first phase has $K$ time
slots. Since the transmitter does not have channel knowledge in this
phase, it sends $N_\textrm{t}=K-1$ independent data symbols using spatial multiplexing on $N_\textrm{t}$ antennas without using any precoding so that each user receives one linear equation for $N_\textrm{t}$ desired symbols while overhearing $K-1$ other transmissions, each of which contains
$K-1$ undesired information symbols. During time slot $n$, the transmitter sends the desired symbols for user
$n$, ${\bf x}[n]={\bf s}^{(n)}$ through $N_\textrm{t}$ antennas. Hence, the received signal at user $k$ in this phase is given by
\begin{align}
{ y}^{(k)}[n]&= {{\bf h}^{(k)T}}[n]{\bf s}^{(n)},\quad n\in\{1,2\ldots,K\},
\end{align}
where $k\in\{1,2\ldots,K\}$ and the noise term was dropped for simplicity because it does not affect
the DoF calculation. 
%
%

\subsubsection{Phase Two}

The second phase spans $K-2$ time
slots. In this phase, the
transmitter has CSI for the current and past time periods because each user sends back both outdated and current CSI from the $K+1$ time slots in phase one.

The main objective of phase two is for the transmitter to provide  each user with additional observations that can be used to build additional $K-2$ linearly
independent equations in the desired symbols. Since each user
has one observation from phase one containing its $K-1$ symbols,
if it obtains additional $K-2$ linearly independent equations
from this phase, then $N_\textrm{t}=K-1$ desired symbols can be decoded at each user. To deliver $N_\textrm{t}$ desired symbols to each user through phase two, the transmitted signal at time slot $n$ is given by
\begin{align}
{\bf x}[n]=\sum_{k=1}^{K}{\bf V}^{(k)}[n]{\bf s}^{(k)}, \label{eq:STIAphase2send}
\end{align}
where $n\in\{K+1,K+2,\ldots,2K-2\}$.
Notice that the transmitter repeatedly
sends the same symbol vector ${\bf s}^{(k)}$ regardless of time index $n$, but using a different beamforming matrix ${\bf V}^{(k)}[n]$, which varies according to time index. When the transmitter sends the signal in (\ref{eq:STIAphase2send}) at
time slot $n$, the received signal at user $\ell$ is given by
\begin{align}
{ y}^{(\ell)}[n]&= {{\bf h}^{(\ell)T}}[n]\sum_{k=1}^{K}{\bf
V}^{(k)}[n]{\bf s}^{(k)} \nonumber \\
&= \underbrace{{{\bf h}^{(\ell)T}}[n]{\bf V}^{(\ell)}[n]{\bf
s}^{(\ell)}}_{{ L}^{(\ell,\ell)}[n]}\!+\!\!\!\sum_{k:k\neq \ell}^{K}\underbrace{{{\bf
h}^{(\ell)T}}[n]{\bf V}^{(k)}[n]{\bf s}^{(k)}}_{{ L}^{(\ell,k)}[n]},
\end{align}
where $\ell\in\{1,2\ldots,K\}$.
Since the transmitter sends a linear combination of $K(K-1)$ data
symbols for all the users in each time slot, each user sees not only
one desired term i.e.,
${ L}^{(\ell,\ell)}[n]$ but
also receives $K-1$ interference terms, i.e., ${ L}^{(\ell,k)}[n]$ for $k\in\{1,2,\ldots,K\}/\{\ell\}$. To decode the $K-1$ desired symbols reliably in high SNR, user $\ell$ should be able to remove the interference equations ${L}^{(\ell,k)}[n]$ for
$k\in\{1,2,\ldots,K\}/\{\ell\}$.

Recall that during the first phase,
user $\ell$ overheard the interfering symbols for user $k$, ${\bf
s}^{(k)}$, at the $k$-th time slot in the form of ${{\bf
h}^{(\ell)T}}[k]{\bf s}^{(k)}=L^{(\ell,k)}[k]$. Therefore, if user $\ell$ sees
the same linear combinations of the interfering symbols ${\bf
s}^{(k)}$ at the $n$-th time repeatedly, i.e., ${{\bf h}^{(\ell)T}}[n]{\bf
V}^{(k)}[n]{\bf s}^{(k)}={{\bf h}^{(\ell)T}}[k]{\bf s}^{(k)}$, then it can remove this interference by subtracting the previously observed signal
as side-information. Therefore, it is important to design the beamforming matrices 
so that the users see the same interference in the first and second phases. To accomplish this goal, the beamforming matrices
should satisfy the space-time interference alignment
condition:
\begin{align}
\underbrace{\left[%
\begin{array}{c}
  {{\bf h}^{(1)T}}[k] \\
  \vdots \\
  {{\bf h}^{(k-1)T}}[k]  \\
  {{\bf h}^{(k+1)T}}[k] \\
  \vdots \\
  {{\bf h}^{(K)T}}[k] \\
\end{array}%
\right]}_{(K-1) \times N_\textrm{t}}&=\underbrace{\left[%
\begin{array}{c}
  {{\bf h}^{(1)T}}[n]  \\
  \vdots \\
  {{\bf h}^{(k-1)T}}[n]   \\
  {{\bf h}^{(k+1)T}}[n]  \\
  \vdots \\
  {{\bf h}^{(K)T}}[n] \\
\end{array}%
\right]}_{ (K-1) \times N_\textrm{t}}\underbrace{{\bf V}^{(k)}[n]}_{N_\textrm{t} \times
N_\textrm{t}},
\\
\Rightarrow {\bf {H}}_k^c[k]&={\bf {H}}_k^c[n]{\bf V}^{(k)}[n],\label{STIA_cond}
\end{align}
where $k\in\{1,2,\ldots,K\}$, $n\in\{K\!+\!1,K\!+\!2,\ldots,2K-2\}$
and, ${\bf {H}}_k^c[n]$ and ${\bf {H}}_k^c[k]$ represent an augmented channel matrix of all users excepting user $k$ at time slot
$n$ and $k$, respectively. This inter-user interference
alignment condition implies that the signal for user $k$, which
interferes with other users excluding user $k$ during the second
phase, should be the same as the interference signal overheard by the other users in time slot $k$ in the first phase. In other words, the transmitter
delivers ${\bf s}^{(k)}$ using the
beamforming matrix ${\bf V}^{(k)}[n]$ in such a way that the other
interfering receivers (excluding user $k$) see the same linear
combination of the interfering data ${\bf s}^{(k)}$, which they
previously observed at time slot $k$ in the first phase. Notice that the rank of ${\bf {H}}_k^c[n]$ becomes $K-1$ with
probability one because $N_\textrm{t}=K-1$ and all channel values are selected from a continuous distribution. Using this fact, the beamforming matrix for user $k$,
${\bf V}^{(k)}\in \mathbb{C}^{N_\textrm{t} \times N_\textrm{t}}$, satisfying the interference alignment condition in (\ref{STIA_cond}) is constructed as \begin{align}
{{\bf {V}}^{(k)}}[n]={{\bf {H}}_k^c[n]}^{-1} {\bf {H}}_k^c[k].
\label{eq:STIA_BF}
\end{align}

Now, we consider decodability achieved by using the designed ${{\bf {V}}^{(k)}}[n]$. To make the exposition concrete, we focus on the decoding process at user 1.
The concatenated received signals at user $1$ during
the both phases are given by
\begin{align}
\left[\!\!\!%
\begin{array}{c}
  { y}^{(1)}[1] \\
  { y}^{(1)}[2] \\
  \vdots \\
  { y}^{(1)}[K] \\ \hline
  { y}^{(1)}[K\!+\!1] \\
  { y}^{(1)}[K\!+\!2] \\
  \vdots \\
  { y}^{(1)}[2K\!-2] \\
\end{array}%
\!\!\right]\!\!=\!\!
\left[\!\!%
\begin{array}{c}
  {{\bf h}^{(1)T}}[1]{\bf s}^{(1)} \\
  {{\bf h}^{(1)T}}[2]{\bf s}^{(2)} \\
  \vdots \\
  {{\bf h}^{(1)T}}[K]{\bf s}^{(k)} \\ \hline
  {{\bf h}^{(1)T}}[K\!+\!1]\sum_{k=1}^{K}{{\bf {V}}^{(k)}}[K+1]{\bf s}^{(k)} \\
  {{\bf h}^{(1)T}}[K\!+\!2]\sum_{k=1}^{K}{{\bf {V}}^{(k)}}[K+2]{\bf s}^{(k)} \\
  \vdots \\
  {{\bf h}^{(1)T}}[2K\!-\!2]\sum_{k=1}^{K}{{\bf {V}}^{(k)}}[2K-2]{\bf s}^{(k)} \\ 
\end{array}%
\!\!\right].
\end{align}
Since we designed the transmit beamforming matrix ${{\bf
{V}}^{(k)}}[n]$ so that it satisfies the space-time
interference alignment conditions in (\ref{eq:STIA_BF}), it is
possible for user 1 to eliminate interference signals received
during the second phase using the overheard equations during the
first phase as side-information. For example, at the $\ell$-th time slot for $\ell\in\{K+1,K+2,\ldots,2K-2\}$, user 1 receives ${\bf h}^{(1)T}[\ell]\sum_{k=2}^{K}{\bf {V}}^{(k)}[\ell]{\bf s}^{(k)}$ interference signal. Since user 1 already received the same interference signals from time slot 2 to time slot $K$, i.e., $\left\{{\bf h}^{(1)T}[2]{\bf s}^{(2)},{\bf h}^{(1)T}[3]{\bf s}^{(3)},\ldots, {\bf h}^{(1)T}[K]{\bf s}^{(K)}\right\}$, it can remove all the interference signals observed in time slot $\ell$.
After interference cancellation, the effective received
signal at user 1 is given by
\begin{small}
\begin{align}
\left[\!\!\!\!%
\begin{array}{c}
  {y}^{(1)}[1] \\ 
  { y}^{(1)}[K\!+\!1]\!-\!\sum_{k=2}^{K}{y}^{(1)}[k] \\
  { y}^{(1)}[K\!+\!2]\!-\!\sum_{k=2}^{K}{ y}^{(1)}[k] \\
  \vdots \\
  { y}^{(1)}[2K-2]\!-\!\sum_{k=2}^{K}{ y}^{(1)}[k] \\
\end{array}\!\!\!\!
\right]&\!\!\!=\!\!\! \underbrace{\left[\!\!\!\!
\begin{array}{c}
  {{\bf h}^{(1)T}}[1] \\ 
  {{\bf h}^{(1)T}}[K+1]{{\bf {V}}^{(1)}}[K\!+\!1]\\
  {{\bf h}^{(1)T}}[K+2]{{\bf {V}}^{(1)}}[K\!+\!2]\\
  \vdots \\
  {{\bf h}^{(1)T}}[2K-2]{{\bf {V}}^{(1)}}[2K\!-\!2]\\
\end{array}\!\!\!%
\right]}_{N_\textrm{t}\times N_\textrm{t}}\!\!{\bf s}^{(1)},\\
\Rightarrow {\bf \hat{y}}^{(1)}&={\bf {\hat{H}}}^{(1)}{\bf
s}^{(1)},
\end{align}
\end{small}
where ${\bf \hat{y}}^{(1)}$ and ${\bf \hat{H}}^{(1)}$ denote the effective received signal vector and channel matrix
at user 1 after interference cancellation. The effective channel matrix
for ${\bf {\hat{H}}}^{(1)}$ has a full rank of
$ N_\textrm{t}$ almost surely because ${\bf
{V}}^{(1)}[K+1]$, ${\bf {V}}^{(1)}[K+2]$, $\ldots$, and ${\bf
{V}}^{(k)}[2K-2]$ are
independently constructed with respect to ${{\bf
h}^{(1)T}}[1]$, ${{\bf h}^{(1)T}}[K+1]$, ${{\bf
h}^{(1)T}}[K+2]$, $\ldots$, and ${{\bf
h}^{(1)T}}[2K-2]$.
Therefore, in the high \textrm{SNR} regime, the receiver can  decode the desired $N_\textrm{t}$ symbols
$s^{(1)}_k$ for $k\in\{1,2\ldots,N_\textrm{t}\}$ using a ZF receiver. By the same argument, it is possible to obtain $N_\textrm{t}=K-1$ data symbols over
$2K-2$ time slots when $\omega=\frac{K-2}{2K-2}$ for the other users as well. As a result, 
$d(\frac{K-2}{2K-2})=\frac{K(K-1)}{2K-2}=\frac{K}{2}$ DoF are achieved
using periodic feedback.

\subsection{Achievability of Point C}
Consider point C where the non-CSI feedback duration equals to one time slot, i.e., $T_{\textrm{n}}=1$, and the CSI feedback duration comprises of $K-1$ time slots, i.e., $T_{\textrm{f}}=K-1$. The achievability proof for point C is also shown by the STIA method.
 
\subsubsection{Phase One}
 Only one time slot is used in phase one. In this time slot, the transmitter sends $N_\textrm{t}=K-1$ independent symbols to their corresponding users at the same time. The main objective in this phase is for each user to see an equation in the form of a superposition of all transmitted data symbols, which will be used as side-information for decoding. The transmitted signal in time slot $1$ is given by
\begin{align}
{\bf x}[1]= \sum_{k=1}^{K}{\bf s}^{(k)},
\end{align}1
where ${\bf
s}^{(k)}=\left[s^{(k)}_1, s^{(k)}_2, \ldots, s^{(k)}_{K-1}\right]^{T}$.
The received signal at user $j$ is given by
\begin{align}
y^{(j)}[1]&= {\bf h}^{(k)T}[1]\sum_{k=1}^{K}{\bf s}^{(k)} \nonumber \\
&={ L}^{(j,j)}[1]+\sum_{k=1,k\neq j}^{K}{ L}^{(j,k)}[1],
\end{align}
where noise is neglected for the DoF analysis as before.

\subsubsection{Phase Two} The remaining $K-1$ time slots are used in phase two. The objective in phase two is for the transmitter to send the information signal in such a way that each user sees the same interference shape observed at time slot $1$. To accomplish this goal, the transmit precoder for carrying the data symbols for user $k$ at time slot $n\in \{2,\ldots, K\}$, is computed as
\begin{align}
{\bf V}^{(k)}[n]&= \left[%
\begin{array}{c}
  {{\bf h}^{(1)T}}[n] \\
\vdots \\
  {{\bf h}^{(k-1)T}}[n] \\
  {{\bf h}^{(k+1)T}}[n] \\
  \vdots\\
  {{\bf h}^{(K)T}}[n] \\
\end{array}%
\right]^{-1}\left[%
\begin{array}{c}
  {{\bf h}^{(1)T}}[1] \\
\vdots \\
  {{\bf h}^{(k-1)T}}[1] \\
  {{\bf h}^{(k+1)T}}[1] \\
  \vdots\\
  {{\bf h}^{(K)T}}[1] \\
\end{array}%
\right], \label{eq:STIA_BF_2}
\end{align}
where $k\in\{1,2\ldots,K\}$. Effectively, the current channel is equalized and forced to mimic the channel corresponding to the old CSI available in time slot 1. The received signal at user $k$ at time slot $n$ is given by
\begin{align}
{ y}^{(k)}[n]&= {\bf h}^{(k)T}[n]\sum_{j=1}^{K}{\bf V}^{(j)}[n]{\bf s}^{(j)}\nonumber \\
&={ L}^{(k,k)}[n]+\sum_{j\neq k}^{K}{ L}^{(k,j)}[n] \nonumber \\
&\stackrel{(a)}{=}{ L}^{(k,k)}[n]+\sum_{j\neq k}^{K}{ L}^{(k,j)}[1],
\end{align}
where (a) comes from the fact that during the second phase, i.e., $n\in\{2,3,\ldots,K\}$, user $k$ observes the same interference pattern $\sum_{j\neq k}^{K}{ L}^{(k,j)}[n]=\sum_{j\neq k}^{K}{ L}^{(k,j)}[1]$ for $n\in\{2,\ldots,K\}$.

\subsubsection{Interference Cancellation} Because of the special choice of precoding matrix in (\ref{eq:STIA_BF_2}), the receiver is able to subtract interference received from the received signal during the second phase using the saved equation from the first phase. Using the received signal in each time slot of phase 2, the receiver constructs 
\begin{align}
 {y}^{(k)}[n]- { y}^{(k)}[1] &= {\bf h}^{(k)T}[n]\sum_{j=1}^{K}{\bf V}^{(j)}[n]{\bf s}^{(j)}\nonumber \\
&-{\bf h}^{(k)T}[1]\sum_{j=1}^{K}{\bf V}^{(j)}[1]{\bf s}^{(j)} \nonumber \\
&={ L}^{(k,k)}[n]+\sum_{j=1,j\neq k}^{K}{ L}^{(k,j)}[n] \nonumber \\
&-{ L}^{(k,k)}[1]-\sum_{j=1,j\neq k}^{K}{ L}^{(k,j)}[1]
\nonumber \\
&={ L}^{(k,k)}[n]-{ L}^{(k,k)}[1], \quad n\in\{2,3,\ldots,K\}.
\end{align}
Stacking all $K-1$ equations together into a linear system gives
 \begin{align}
\!\!\!\!\!\left[\!\!\!%
\begin{array}{c}
{y}^{(k)}[2]- { y}^{(k)}[1]\\
{ y}^{(k)}[3]- { y}^{(k)}[1]\\
\vdots \\
{ y}^{(k)}[K]- { y}^{(k)}[1]\\
\end{array}\!\!\!%
\right]\!\!\!&=\!\!\!\left[
\!\!\begin{array}{c}
{ L}^{(k,k)}[2]-{ L}^{(k,k)}[1]\\
{ L}^{(k,k)}[3]-{ L}^{(k,k)}[1]\\
\vdots \\
{ L}^{(k,k)}[K]-{ L}^{(k,k)}[1]\\
\end{array}%
\right] \nonumber \\
&\!\!\!=\!\!\left[\!\!\!%
\begin{array}{c}
{\bf h}^{(k)T}[2]{\bf V}^{(k)}[2]-{\bf h}^{(k)T}[1]\\
{\bf h}^{(k)T}[3]{\bf V}^{(k)}[3]-{\bf h}^{(k)T}[1]\\
\vdots \\
{\bf h}^{(k)T}[K]{\bf V}^{(k)}[K]-{\bf h}^{(k)T}[1]\\
\end{array}%
\!\!\!\right] {\bf s}^{(k)} \nonumber \\
&=\!{\bf \bar{H}}^{(k)}{\bf s}^{(k)},
\end{align}
where $k\in\{1,2,\ldots,K\}$. Now recall the following facts. First, the precoding matrices ${\bf V}^{(k)}[n]$ in (\ref{eq:STIA_BF_2}) are independently generated regardless of the direct channel ${\bf h}^{(k)T}[n]$ for $n\in\{2,3\ldots,K\}$. Second, the channel elements come from a continuous random distribution. Using these two facts, it follows that ${\bf h}^{(k)T}[n]{\bf V}^{(k)}[n]$ are statistically independent. Further, since ${\bf h}^{(k)T}[n]{\bf V}^{(k)}[n]$ and ${\bf h}^{(k)T}[1]$ are linear independent for $\forall n$ and $\forall k$, it follows that $\textrm{rank}\left({\bf \bar{H}}^{(k)}\right)=K-1$ almost surely. Therefore, by applying a ZF decoder, user $k$ obtains ${\bf s}^{(k)}$. Consequently, for the $K$-user $(K-1) \times 1$ MISO broadcast channel, each transmitter delivers $(K-1)$ data symbols to the corresponding user over $K$ time slots when $\omega=\frac{K-1}{K}$, which leads to achieve $d\left(\frac{K-1}{K}\right)=K-1$ DoF. \endproof

\textbf{Remark 4 (Least Squares STIA)}: It is possible to extend the STIA algorithm to $N_\textrm{t} < (K-1)$, at the expense of DoF. As shown in (\ref{eq:STIA_BF}) and  (\ref{eq:STIA_BF_2}), perfect STIA is accomplished when $N_\textrm{t}\geq (K-1)$. If $N_\textrm{t} < (K-1)$, the matrix in (\ref{eq:STIA_BF_2}) is not invertible. Consequently, we propose to find the precoder that minimizes but not necessarily eliminates the residual interference in the alignment condition. By minimizing the error residual in the interference alignment condition, $\min_{{\bf V}^{(k)}[n]} \parallel{\bf {H}}_k^c[k]-{\bf {H}}_k^c[n]{\bf V}^{(k)}[n]\parallel_{F}$, the least squares STIA (LS-STIA) is obtained as
\begin{align}
{{\bf \hat{V}}^{(k)}_{LS}}[n]=\left({\bf
{H}}_k^c[n]^{*}{{\bf {H}}_k^c[n]}\right)^{-1}{{\bf {H}}_k^c[n]}^{*}{\bf {H}}_k^c[k].
\label{eq:LS_STIA_BF}
\end{align}
Although LS-STIA solution in (\ref{eq:LS_STIA_BF}) does not achieve the same DoF performance compared to the perfect STIA in (\ref{eq:STIA_BF}) and (\ref{eq:STIA_BF_2}), it could be a robust solution to the alignment error when $N_\textrm{t} < (K-1)$. Precisely quantifying the DoF degradation and practical achievable rates is a topic for future work.


\subsection{Analysis of the DoF Gain from Outdated CSI Feedback}
This section quantifies the DoF gain due to outdated CSI feedback relative to a baseline strategy when each user does not feedback CSI to the transmitter during the non-feedback period $T_{\textrm{n}}$, and sends back current CSI only during the feedback period $T_{\textrm{f}}$. The baseline transmission strategy is to select alternatively a transmission method between the ZF method when CSIT is available and the TDMA method when CSIT is not available. The achievable sum-DoF region $d_{\textrm{ZF-TDMA}}(\omega)$ by this baseline transmission strategy is given by
\begin{align}
d_{\textrm{ZF-TDMA}}(\omega)&=(K-1)\omega + (1-\omega) \nonumber\\
&= (K-2)\omega +1 \quad \textrm{for} \quad \omega\in[0,1].  
\end{align}

By computing the different between $d(\omega)$ in Theorem 1 and $d_{\textrm{ZF-TDMA}}(\omega)$, we observe how much DoF gain we can obtain from additional outdated CSI feedback in the periodic CSI feedback as shown in Fig. 3. There are two interesting observations. One is that outdated CSI feedback always improves the DoF gain over all feedback frequency in the range $0<\omega<1$. Further, when feedback the frequency is larger or equal to $\frac{K-1}{K}$, outdated CSI feedback does not degrade the DoF gain performance. This DoF gain comes from joint exploitation of current and outdated CSI, which allows each receiver to exploit its overheard side-information to achieve DoF gains compared to the baseline case.

\subsection{Comprison with an Outer Bound \cite{Tandon_out}}
Recently, a DoF outer bound result was established in \cite{Tandon_out} for the alternating CSIT model \cite{Tandon_alt} in the vector broadcast channel. Interestingly, it was shown in \cite{Tandon_out} that the transmitter requires to use perfect CSI during the $\min\{N_\textrm{t},K\}/K$ fraction of time in minimunm to achieve the $\min\{N_\textrm{t},K\}$ cut-set DoF by a new converse proof technique. From this converse argument, the proposed region in Theorem 1 has  the DoF optimality for $\omega\in[\frac{K-1}{K}, 1]$ and $\omega =0$. For $0<\omega < \frac{K-1}{K}$, however, the DoF gap between the outer bound result in \cite{Tandon_out} and the proposed region exists as illustrated in Fig. 3. Therefore, finding the optimal sum-DoF region over the range of $0<\omega < \frac{K-1}{K}$ is still an open problem. Further, one interesting open problem is to investigate the DoF region in the case when $N_{\textrm{t}}<K-1$. We conjecture that increasing the number of users would help to decrease the feedback frequency requirement from the observation in the outer bound result \cite{Tandon_out}. As shown in Remark 4, however, the proposed STIA algorithm does not satisfy the perfect alignment condition for the $N_{\textrm{t}}<K-1$ case due to a feasibility condition issue leading to DoF loss. Therefore, to show tightness of this outer bound in the periodic CSI feedback framework, a new achievable scheme would be needed for the $N_{\textrm{t}}<K-1$ case.

\section{CSI Feedback Delay-DoF Gain Trade-Off}
In this section, we characterize the achievable sum-DoF region for the periodic CSI feedback Model 2 in Section II. We quantify the relationship between the CSI feedback delay and the sum-DoF gain for the vector broadcast channel leveraging the concept of STIA and using the results in \cite{Maddah-Ali2}. Through this trade-off analysis, we provide insights into the interplay between CSI feedback delay and system performance from a sum-DoF gain perspective.

We devote to prove the following theorem in this section.

\begin{theorem} \label{Theorem 2}
\emph { For the $K$-user $(K-1) \times 1$ MISO broadcast channel with periodic CSI feedback (Model 2), the achievable trade-off region between CSI feedback delay and the sum-DoF gain is given by
\begin{align}
d(\gamma) \!=\!\! \left\{\!\!
\begin{array}{l l}
  K-1, & \!\!\!\!\!\!\quad \textrm{for} \quad 0\leq \gamma\leq \frac{1}{K}, \\
  \left(\!-K\!+\!\frac{Kc(K)}{K-1}\!\right)\!\gamma\!+\!K\!-\!\frac{c(K)}{K\!-\!1}, & \!\!\!\!\!\!\quad \textrm{for} \quad \frac{1}{K}< \gamma\leq 1,\\
 c(K), & \!\!\!\!\!\!\quad \textrm{for} \quad \gamma\geq 1.\\
\end{array} \right.  
\end{align} where $c(K)= \frac{K-1}{1+\frac{1}{2}+\cdots+\frac{1}{K-1}}$.}
\end{theorem}

\subsection{Achievability Proof for Theorem 2}

The main technical challenge in this proof is to show that $d(\frac{1}{K})=\min\{N_\textrm{t},K\}=K-1$. The other region $d(\gamma)=K-1$ for $ 0 < \gamma \leq\frac{1}{K}$ follows from this result using time sharing between the proposed STIA and ZF. Further, for the region of $ \frac{1}{K} < \gamma \leq 1 $, achievability of $d(\gamma)=\left\{-K+\frac{Kc(K)}{K-1}\right\}\gamma+K-\frac{c(K)}{K-1}$ can be shown by using time sharing between the proposed STIA and the Maddah-Ali-Tse (MAT) method in \cite{Maddah-Ali2}. 

\begin{figure}
\centering
\includegraphics[width=3.5in]{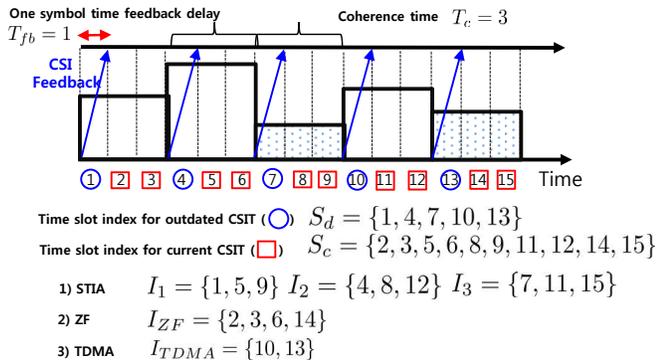}
\caption{Periodic CSI feedback (Model 2) when $n=3$, $K=3$, and $\gamma=\frac{1}{3}.$} \label{fig:resource}
\end{figure}

Without loss of generality, we assume that the channel coherence $T_{\textrm{c}}$ has $K$ time slots and the feedback delay time is just one time slot $T_{\textrm{fb}}=1$ so that $\gamma=\frac{1}{K}$. Under this assumption, if the user sends back CSI at the first time slot of the $t$-th channel block ${t}_1$, then the transmitter can have current CSI at $\{{t}_2,{t}_3,\ldots,{t}_K\}$ time slots due to one time slot feedback delay. Under this channel knowledge assumption, we show that $d(\frac{1}{K})=N_\textrm{t}$ of DoF are achievable asymptotically if $N_\textrm{t}=K-1$. Consider $n+K-1$ channel blocks, which are comprised of a total of $Kn+K(K-1)$ time slots so that $S_t=\{1,2,3,4,\ldots,Kn+K(K-1)\}$ for transmission. Since we assume that the normalized feedback delay is $\frac{1}{K}$, the total time slot set can be decomposed two subsets, $S_c$ with $|S_c|=(K-1)(n+K-1)$ and $S_d$ with $|S_d|=n+K-1$. Here, $S_c$ and $S_d$ represent time slot sets corresponding to the case where the transmitter has current and delayed CSI. Now, define $n$ time slot sets, $\left\{I_{1},I_2,\ldots, I_n\right\}$, each of which has $K$ elements for STIA algorithm, i.e., $I_{\ell}=\{k_1,k_2,\ldots,k_K\}$ where $k_1\in S_{d}$, $k_j\in S_c$ for $j\in\{2,3,\ldots,K\}$, and $k_{\ell}$ and $k_j$ belong to a different channel coherence time block when $\ell \neq j$. Further, let us define two different time slot sets  $I_{\textrm{ZF}}$  for ZF and $I_{\textrm{TDMA}}$ for TDMA transmission. Since the used time slots for STIA transmission are $|I_1\cup I_2 \cup \cdots \cup I_n|=Kn$, the time slot set for ZF and TDMA is defined as
\begin{align}
I_{\textrm{ZF}}=S_c - (I_1\cup I_2 \cup \cdots \cup I_n) \nonumber \\
I_{\textrm{TDMA}}=S_d - (I_1\cup I_2 \cup \cdots \cup I_n), \nonumber
\end{align}
where $|I_{\textrm{ZF}}|=(K-1)^2$ and $|I_{\textrm{TDMA}}|=K-1$. For example, as shown in Fig. \ref{fig:resource}, when $K=3$ and $n=3$, there are a total of $3n+6=15$ time resources. The temporal resources can be partitioned into five indexed sets defined as $I_1=\{1,5,9\}$, $I_2=\{4,8,12\}$, $I_3=\{7,11,15\}$, $I_{\textrm{ZF}}=\{2,3,6,14\}$, and $I_{\textrm{TDMA}}=\{10,13\}$. The key idea in this proof is to apply different transmission strategies: the proposed STIA, ZF and TDMA  according to different subsets of time
slots.

\subsubsection{STIA}
We first show that $nK(K-1)$ DoF are achievable using the proposed STIA algorithm when $nK$ time slots are used  i.e, $|I_1\cup I_2 \cup \cdots \cup I_n|=Kn$.
This proof is identical to the achievability proof for Point C in Theorem 1, which was shown in Section IV. Since for given $I_{\ell}$, the transmitter is able to use $K-1$ current and outdated CSI over a total $K$ slots, the normalized delay parameter $\gamma=\frac{1}{K}$ is equivalently interpreted in terms of the feedback frequency $\omega$ with the relationship of $\omega=\frac{K-1}{K}=1-\gamma$. From this equivalence, we conclude that $K(K-1)$ DoF are achievable by spending each time slot set $I_{\ell}$ with $K$ elements for $\ell\in\{1,2,\ldots,n\}$.

\subsubsection{TDMA and ZF}
Since $Kn$ time slots have been used for STIA among the total temporal resources$Kn+K(K-1)$, $I_{\textrm{ZF}}$ and $I_{\textrm{TDMA}}$ time
slot sets remain where $|I_{\textrm{ZF}}|=(K-1)^2$ and $|I_{\textrm{TDMA}}|=K-1$. Here, we use an alternative transmission method depending on CSIT knowledge. The transmitter sends data by using TDMA because CSIT is not available due to feedback delay. Alternatively, the transmitter delivers multiple data streams by using ZF because the transmitter is able to use CSI. From this method, the achievable DoF is $d(\frac{1}{K})=(K-1)(K-1)^2+ (K-1)$ in the remaining $K(K-1)$ time slots. 

\subsubsection{Asymptotic the Sum-DoF Gain} 
We have divided the total time resource $S_t$ with $|S_t|=Kn+K(K-1)$ into three different groups and applied a different transmission method in each group. Hence, using time sharing, the total sum-DoF gain is computed as
\begin{align}
d\left(\frac{1}{K}\right)&=\!\frac{\underbrace{(K-1) Kn}_{\textrm{STIA}}+\underbrace{ (K-1) (K-1)^2}_{\textrm{ZF}}+\underbrace{ (K-1)}_{\textrm{TDMA}}  }{Kn+K(K-1)} \nonumber\\
&=\frac{(K-1)Kn+(K-1)^3+(K-1)}{Kn+K(K-1)}.
\end{align}
Therefore, for a given $K$, as $n$ goes to infinity, the sum-DoF gain asymptotically achieves
\begin{align}
\lim_{n\rightarrow \infty}d\left(\frac{1}{K}\right) =K-1,
\end{align}
which completes the proof.
\endproof

\subsection{Special Case: Three-user Two-Antenna Vector Broadcast Channel}
In this section, we provide an exact characterization of the CSI feedback delay-DoF gain trade-off region for the 3-user $2\times 1$ MISO broadcast channel.
Achievability is shown by the result in Theorem 2 and the enhanced DoF result in Theorem 5 \cite{Maddah-Ali2}. Further, a converse is proven using the result in \cite{Tandon_out} by considering a minor modification for a block fading channel.

\begin{corollary} \label{Corollary1}
The optimal CSI feedback delay-DoF gain trade-off region for the 3-user $2\times 1$ MISO broadcast channel is given by
\begin{align}
d^*(\gamma) = \left\{
\begin{array}{l l}
  2, & \quad \textrm{for} \quad 0\leq \gamma\leq \frac{1}{3}, \\
  -\frac{3}{4}\gamma+\frac{9}{4}, & \quad \textrm{for} \quad \frac{1}{3}< \gamma\leq 1,\\
  \frac{3}{2}, & \quad \textrm{for} \quad \gamma\geq 1.\\
\end{array} \right.\\ \nonumber \label{eq:co1}
\end{align}
\end{corollary}

\subsubsection{Achievability}
Achievability of this corollary can be proved by applying time sharing technique between STIA and the enhanced transmission method proposed in Theorem 5 \cite{Maddah-Ali2}. 

\subsubsection{Converse}
For a proper converse argument, let us consider the following lemma given by \cite{Tandon_out}.
\lemma For the 3-user $2\times 1$ MISO broadcast channel with identical channel coherence patterns and the same normalized feedback delay $\gamma$ across the users, the following DoF bound holds: 
\begin{eqnarray}
d^{(1)} +d^{(2)}+d^{(3)} \leq \frac{9}{4}-\frac{3}{4} \gamma. \label{eq:outer}
\end{eqnarray}
\proof A detailed proof is provided in \cite{Tandon_out} for the fast fading scenario where the channel gains change independently across time for all users. Here, we consider a minor modification for a block fading channel where all users have identical channel coherence patterns. The idea of the converse proof in \cite{Tandon_out}  is to covert the original 3-user $2\times 1$ MISO broadcast channel into a  2-user $2\times 1$ MISO physically degraded broadcast channel by allowing cooperation between user 2 and user 3 and providing the channel output signal of user 1 with the cooperating user group. Using the fact that feedback does not increase the capacity of the physically degraded broadcast channel and its capacity depends on the marginal distribution only, it is possible to create $\min\{N_{\textrm{t}},K\}-1=1$ artificial receivers that can decode the same message of user 1. By applying the differential entropy bound techniques used in \cite{Tandon_out},  the bounds follows
\begin{eqnarray}
2d^{(1)} +d^{(2)}+d^{(3)} \leq 2 + \alpha,  \label{eq:out1}\\  
d^{(1)} +2d^{(2)}+d^{(3)} \leq 2 + \alpha,  \label{eq:out2}\\ 
d^{(1)} +d^{(2)}+2d^{(3)} \leq 2 + \alpha,  \label{eq:out3}
\end{eqnarray}
where $\alpha$ denotes the fraction of time instances where the perfect CSIT is available. By replacing $\alpha=1-\gamma$ and adding three inequalities in (\ref{eq:out1}), (\ref{eq:out2}), and (\ref{eq:out3}), the inequality of (\ref{eq:outer}) results, completing the proof.
\endproof
Using Lemma 1 and the cut-set bound $d^{(1)} +d^{(2)}+d^{(3)} \leq \min\{K,N_{\textrm{t}}\}=2$, it concludes that the region in (48) is the optimal.

\begin{figure}
\centering
\includegraphics[width=3.5in]{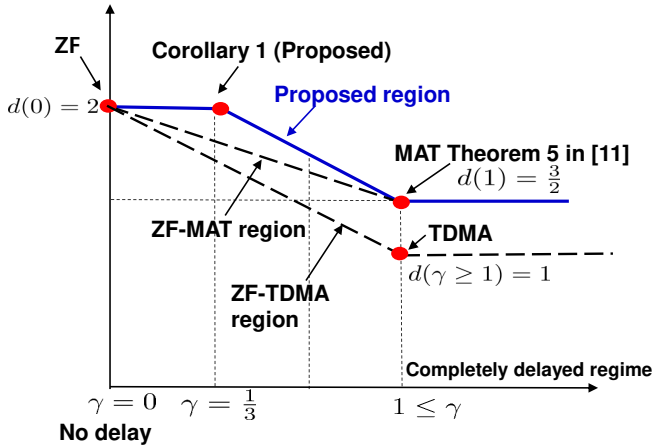}
\caption{The optimal CSI feedback delay-DoF gain trade-off for the 3-user
$2\times 1$ MISO broadcast channel.} \label{fig:corollary1}
\end{figure}
\begin{table*}
\begin{center}
  \caption{Comparison
of three different interference alignment algorithms}\small 
\begin{tabular}{c|c|c|c} \hline
   Scheme &  Fading type  & Required CSIT  & Feed-forward overhead   \\ \hline\hline
  Blind IA \cite{JafarBIA}  & Staggered block fading  &  Fading autocorrelation function &  No \\ \cline{1-4}
  MAT  &  General fading & Outdated CSI &  CSIR for other receivers   \\ \cline{1-4}  
  STIA  & General fading & Outdated and current CSI & No \\\hline
\end{tabular}\label{Table2}
\end{center}
\end{table*}

\textbf{Remark 5 (Sum-DoF Region Comparison)}: We will compute the achievable CSI feedback delay-DoF gain tradeoff curves achieved by time sharing among conventional schemes, and compare them against the proposed optimal tradeoff curve. By time sharing between ZF and TDMA, it is possible to show that the $d_{\textrm{ZF-TDMA}}(\gamma)=-\gamma+2$ curve is achievable for $0\leq \gamma \leq 1$. Similarly, if we consider the time sharing method between ZF and MAT, then the $d_{\textrm{ZF-MAT}}(\gamma)= -\frac{1}{2}\gamma +2$ of curve is achieved for $0\leq \gamma \leq 1$. As illustrated in Fig. \ref{fig:corollary1}, time sharing between the proposed transmission method and the MAT method in Theorem 5 \cite{Maddah-Ali2} achieves a higher CSI feedback delay-DoF trade-off region. For instance, the proposed algorithm achieves the 2 of DoF when $\gamma=\frac{1}{3}$, which is the $\frac{1}{3}$ of DoF gain over ZF-TDMA and $\frac{1}{6}$ of DoF gain over ZF-MAT.

\textbf{Remark 6}: The proposed CSI feedback delay-DoF gain trade-off shows that if users send CSI to the transmitter within 33$\%$ of the channel coherence time, then the system performance is not degraded from a DoF perspective. Based on previous work \cite{Yang}-\cite{Andrews}, it was understood that even a small amount of CSI feedback delay resulted in degraded DoF. Our result confirms the intuition that a  delay corresponding to a  reasonable fraction of the coherence time does not change the sum-DoF scaling. We are able to precisely quantify how much delay is allowed - finding that it corresponds to one third of the channel coherence time.

\textbf{Example}: Consider a LTE system using $f=2.1$ GHz carrier frequency, which serves users with mobility of $v=3$ km/h (walking speed). In this case, the channel coherence time is roughly calculated as $T_{\textrm{c}} \simeq \frac{c}{8fv}= 21.4$ msec (two radio frames) where $c$ denotes the speed of light. Therefore, if the users can feedback CSI within 7.133 msec (7 subframes), then there is no performance loss from a sum-DoF point-of-view. Based on this observation, the proposed STIA algorithm can be interpreted as a CSI delay robust transmission algorithm.

\section{Comparisons with Similar Interference Alignment Techniques}
In this section, we compare the proposed transmission
method with the blind interference alignment \cite{JafarBIA} and the MAT method in \cite{Maddah-Ali2}. Our STIA algorithm has similarities with blind interference
alignment \cite{JafarBIA}, \cite{Wang} and the MAT algorithm
\cite{Maddah-Ali2}, \cite{Maleki}. During phase one, none of
the transmission algorithms require CSI and simply send multiple
data streams without preprocessing. The only difference occurs in the second phase. Each algorithm performs alignment in the second phase in a slightly different way. Blind interference
alignment exploits a special channel coherence structure called staggered block fading at the transmitter. To realize blind interference alignment, each user may feedback the fading autocorrelation function of its channel. Using blind interference alignment with receive antenna switching, it was shown that
$\frac{N_{\textrm{t}}K}{N_{\textrm{t}}+K-1}$ DoF gain is
achieved in \cite{Wang}. Therefore, for $N_{\textrm{t}}=K-1$ and $N_{\textrm{r}}=1$ case, blind interference alignment and the proposed STIA both achieve the
same DoF gain of $\frac{K(K-1)}{2K-2}=\frac{K}{2}$ even if the
channel knowledge assumption and fading scenarios are different.
 
Unlike blind interference alignment, the MAT algorithm realizes the alignment in the second phase by using delayed CSI under a general i.i.d. fading channel assumption. The basic idea of the MAT algorithm is to exploit the
outdated CSI to swap the received equations that they have
already seen during the first phase between users. By swapping the received equations, each user decodes the desired data symbols while eliminating the aligned interference signal.
To apply the MAT algorithm, each user should feedback CSI whenever their channel changes so that the transmitter can track all the channel variations. Further, each receiver needs to know the CSI of other
receivers since the newly obtained linear combination of the
desired symbols is obtained by swapping during the second
phase. Therefore, to learn the CSI of other receivers, the
transmitter must forward the CSI of other receivers to each user, which
may cause additional overhead.

The proposed STIA exploits both outdated and current CSI offered by the periodic feedback mechanism under a general i.i.d. fading channel assumption. Due to the requirement for current CSI at some instant, it places a more restrictive feedback delay requirement compared to the MAT algorithm \cite{Maddah-Ali2}. Also, it incurs the same CSI feedback overhead as the MAT method because CSI must be sent back to the transmitter when it changes. The proposed algorithm, however, avoids the need for additional feed-forward overhead. This reason is that the effective channel can be estimated using the demodulated reference signal as explained in Remark 3. We summarize the similarity and differences of the three different
transmission algorithms as in Table \ref{Table2}.

\section{Conclusion} \label{sec:Conclusions}
We proposed a new STIA algorithm that exploits
both current and outdated CSI jointly for the MISO broadcast channel with two different periodic CSI feedback models. For the CSI feedback frequency limited model, we characterized the achievable sum-DoF region as a function of the feedback frequency, and showed that the joint exploitation of current and outdated CSI provides a significant sum-DoF gain compared to the case when the transmitter uses current CSI only. For the CSI feedback link delay limited model, we derived a trade-off region between CSI feedback delay and the sum-DoF gain for the MISO broadcast channel using the proposed STIA algorithm and leveraging results in \cite{Maddah-Ali2}. Based on this trade-off, we provided insights into the interplay between CSI feedback delay and system performance from a DoF gain perspective.

\begin{biographynophoto}
{Namyoon Lee}  is a Ph. D student in the Department of Electrical and Computer Engineering at The University of Texas at Austin. He received his M.Sc. degree in Electrical Engineering from KAIST, Daejeon in 2008 and B.Sc degree in Radio and Communication Engineering from Korea University, Seoul, Korea in 2006. From 2008 to 2011, he was with Samsung Advanced Institute of Technology (SAIT) and Samsung Electronics Co. Ltd. in Korea, where he studied next generation wireless communication networks and involved standardization activities of the 3GPP LTE-A (Long Term Evolution Advanced). He also was graduate intern in advanced network topologies team within the Wireless Communications Lab (WCL) at Intel Labs, Santa Clara, CA, USA in 2012. His primal research interest is to developing and analyzing future wireless communication systems using tools including network information theory and stochastic geometry.

   Mr. Lee was a recipient of the 2009 Samsung Best Paper Award. He also was awarded several fellowships, including the Graduate Student Research Fellowship from the Korea Science and Engineering Foundation (KOSEF) in 2006; the Korea Government Fellowship from 2006 to 2007; the Kwanjeong Educational Foundation Fellowship in 2011-2014; and MCD Fellowship in 2012.
\end{biographynophoto}

\begin{biographynophoto}
{Robert W. Heath, Jr.} is a Professor in the Department of Electrical and Computer Engineering at The University of Texas at Austin and is a member of the Wireless Networking and Communications Group where he directs the Wireless Systems Innovations Laboratory. He received his B.S.E.E. (1996) and his M.S.E.E. (1997) degrees from the University of Virginia, and the Ph.D.E.E. (2002) degree from Stanford University. From 1998 to 1999, he was a Senior Member of the Technical Staff at Iospan Wireless Inc, San Jose, CA where he played a key role in the design and implementation of the physical and link layers of the first commercial MIMO-OFDM communication system. The proposed broadband wireless access system employed OFDM modulation, MIMO technology, and adaptive space-time modulation. From 1999 to 2001 he served as a Senior Consultant for Iospan Wireless Inc. In 2003 he founded MIMO Wireless Inc, a consulting company dedicated to the advancement of MIMO technology. Since January 2002, he has been with the Department of Electrical and Computer Engineering at The University of Texas at Austin where he serves as an Associate Professor and is affiliated with the Wireless Networking and Communications Group. He has consulted for a variety of international companies.

Prof. Heath's current research include all aspects of MIMO communication including antenna design, practical receiver architectures, limited feedback techniques, mobility management, and scheduling algorithms as well as cognitive radio, 60GHz wireless communication, and body area networks. He teaches courses on a variety of topics including Wireless Communication, Digital Signal Processing, Space-Time Communication, and Statistical Signal Processing, to support his research program in wireless communication systems. He has published over 150 refereed conference and journal papers, been awarded 20 patents, and is the author of the forthcoming book Advanced MIMO Communication. Dr. Heath has been an Editor for the IEEE Transactions on Communication and an Associate Editor for the IEEE Transactions on Vehicular Technology. He is a member of the Signal Processing for Communications Technical Committee in the IEEE Signal Processing Society. He was a technical co-chair for the 2007 Fall Vehicular Technology Conference, is the general chair of the 2008 Communication Theory Workshop, is a co-organizer and co-technical-chair of the 2009 Signal Processing for Wireless Communications Workshop, and is co-technical-chair of the 2010 International Symposium on Information Theory. He is the recipient of the David and Doris Lybarger Endowed Faculty Fellowship in Engineering and is a registered Professional Engineer in Texas.
\end{biographynophoto}

\end{document}